\documentclass[aps,prx,twocolumn,groupedaddress,showpacs,preprintnumbers,amsmath]{revtex4-1}
\usepackage{epsfig,amsmath,amssymb,graphics,color,calc,hyperref}
\usepackage{mathtools}

\begin{document}

\title{The L\'evy-Rosenzweig-Porter random matrix ensemble}

\author{G. Biroli\textsuperscript{1,2}, and M. Tarzia\textsuperscript{3,4}}

\affiliation{\textsuperscript{1}\mbox{Laboratoire de Physique de l'Ecole Normale Sup\'erieure, ENS}\\
	\textsuperscript{2}\mbox{Universit\'e PSL, CNRS, Sorbonne
		Universit\'e, Universit\'e de Paris, F-75005 Paris, France}\\
	\textsuperscript{3} \mbox{LPTMC, CNRS-UMR 7600, Sorbonne Universit\'e, 4 Pl. Jussieu, F-75005 Paris, France}
	\textsuperscript{4} \mbox{Institut  Universitaire  de  France,  1  rue  Descartes,  75231  Paris  Cedex  05,  France}}

\begin{abstract}
	In this paper we consider an extension of the Rosenzweig-Porter (RP) model, the L\'evy-RP (L-RP) model, in which the off-diagonal matrix elements  are broadly distributed, providing a more realistic benchmark to develop an effective description of non-ergodic extended (NEE) states  in interacting many-body disordered systems.
	We put forward a simple, general, and intuitive argument that allows one to unveil the multifractal structure of the mini-bands in the local spectrum when hybridization is due to  anomalously large transition amplitudes in the tails of the distribution. The idea is that the energy spreading of the mini-bands can be determined self-consistently by requiring that the maximal hybridization rate ${\cal H}_{ij}$ 
	between a site $i$ and the other  $N^{D_1}$ sites of the support set is of the same order  of the Thouless energy itself $N^{D_1 - 1}$. This argument yields the fractal dimensions that characterize the statistics of 
	the multifractal wave-functions in the NEE phase, as well as the whole phase diagram of the L-RP ensemble. Its predictions  are confirmed both analytically, by a thorough investigation of the self-consistent equation for the local density of states obtained using the cavity approach, and numerically, via extensive exact diagonalizations.
\end{abstract}


\maketitle

\section{Introduction}

The appearance of non-ergodic extended (NEE) eigenstates which are neither localized nor fully ergodic and occupy a sub-extensive part of the whole accessible Hilbert space has emerged as a fundamental property of many physical problems, including Anderson and many-body localization, random matrix theory, quantum information, and even quantum gravity.
Multifractal one-particle wave-functions have been proven to appear exactly at the transition point of the Anderson localization (AL) problem~\cite{wegner,rodriguez}. 
In presence of interactions, although the existence of the many-body localization (MBL) transition~\cite{BAA} is now well established at least for one dimensional systems (see also Ref.~\cite{Gornyi} and  Refs.~\cite{reviewMBL1,reviewMBL2,reviewMBL3,reviewMBL4,reviewMBL5} for recent reviews), the investigation of  NEE phases is far from being completed.
On the one hand, recent numerical results~\cite{mace,alet,laflorencie,war} 
and perturbative calculations~\cite{resonances1,resonances2,tarzia} strongly indicate that the many-body eigenstates are multifractal in the whole insulating regime.
On the other hand, 
a sub-diffusive behavior has been often found in the delocalized phase of such systems preceding the MBL transition~\cite{subdiff1,subdiff2} raising the possibility of the existence of a NEE regime also in the delocalized side of the phase diagram~\cite{dinamica}, as originally suggested in the seminal work of Ref.~\cite{levitov}. Strong indications in favour of such a phase, often nicknamed as ``bad metal'', have been recently reported in the out-of-equilibrium phase diagram of the quantum version of the Derrida's Random Energy Model~\cite{qrem1,qrem2,qrem3,qrem4,qrem5}, which can be thought as the simplest mean-field quantum spin glass.
The hierarchical, multifractal structure of eigenstates and hence of  the local spectrum (fractal mini-bands) in interacting qubit systems is also relevant in the context of quantum computation, since it is believed to play a key role  in the search algorithms based on the efficient population transfer~\cite{boixo}. 
Finally, fractal eigenfunctions  were  recently observed and intensively investigated  in the context of  Josephson junction chains~\cite{jj}, and even in the Sachdev-Ye-Kitaev model of quantum gravity~\cite{syk}.

Although the existence of multifractal eigenstates is of principle importance in physical systems as it implies the breakdown of conventional Boltzmann statistics, 
 the properties of the NEE phases, their analytic description, and the understanding of the physical mechanisms that produce them are still far from being well established.

Inspired by the success of random matrix theory, whose predictions are relevant in such seemingly different fields of physics~\cite{rmt1,rmt2},  Kravtsov {\it et al.} 
proposed a solvable random matrix model~\cite{kravtsov}, the generalised Rosenzweig-Porter (RP) model~\cite{RP}, in which fractal wave-functions appear in an intermediate region of the phase diagram, sandwitched between the fully ergodic and the fully localized phases. The RP model has been intensively investigated over the past few years~\cite{warzel,facoetti,bogomolny,bera,pino,shapiro1,shapiro2}, as it provides a playground to explore the nature and the properties of  NEE states. 
Nonetheless, the RP model is largely oversimplified: Differently from realistic many-body systems, the mini-bands in the local spectrum are fractal and not {\it multi}fractal, the spectrum of fractal dimension is degenerate, and anomalously strong resonances are absent.
In fact, in the RP model every site of the reference space, represented by a matrix index, is connected to every other site with the transition amplitude distributed according to the Gaussian law.
In more realistic interacting models delocalization of the wave-functions is due to a long series of quantum transitions, and the effective transition rates between distant states in the Hilbert space are in general correlated and broadly distributed~\cite{tarzia,war,roy}, due to the appearance of strong far-away resonances.

In order to overcome, at least partially, these issues and to formulate a more realistic effective description of the NEE phase, very recently an extension of the RP model, called the LN-RP ensemble, in which the off-diagonal matrix elements have a wide log-normal distribution, has been introduced and studied~\cite{kravtsov1,khay}. In this paper we consider another very natural generalization of the RP model  with power-law distributed off-diagonal matrix elements, first introduced in Ref.~\cite{monthus-LRP}, which we dub the L\'evy-Rosenzweig-Porter (L-RP) ensemble. 
Differently from the Gaussian RP case, hybridization of the energy levels in the L-RP ensemble can be produced by anomalously large transition amplitudes in the tails of the distribution which  cannot be described by perturbation theory (see App.~\ref{sec:first}). 
	 We present two complementary strategies  which circumvent this difficulty. These strategies are able to take into account the effect of the broadly distributed off-diagonal matrix elements in a self-consistent way and to unveil the  multifractal statistics of the eigenstates.
	The first strategy consists in a very simple and physically intuitive extreme value statistics argument: The idea is that the size of the support set  of the mini-bands in the local spectrum, $N^{D_1}$, can be determined self-consistently by requiring that the largest hopping amplitudes between a site $i$ and the other $N^{D_1}$ sites belonging to the same support set are of the same order of the energy spreading of the mini-band itself, $E_{\rm Th} \propto N^{D_1-1}$.
	The second strategy is based on the cavity equations for the resolvent, which become asymptotically exact in the thermodynamic limit, providing a way to resum the whole perturbative series in a self-consistent way. Within this framework the multifractal statistics can be directly accessed by computing analytically the asymptotic scaling behavior of the typical value of the local density of states (LDoS) in the NEE regime. These two approaches give exactly the same predictions for  the phase diagram 
of the L-RP ensemble as a function of the parameters $\mu$ (which characterize the exponent of the tails of the distribution of the transition amplitudes) and $\gamma$ (which characterize the scaling of their typical value with the system size $N$), as well as for the anomalous dimensions of the eigenstates in the NEE phase.
We complement these results by extensive exact diagonalizations which confirm the theoretical analysis and allows one to investigate in great detail the properties 
of the phase transitions between the ergodic, NEE, and AL phases. Our results for the Thouless energy (and thus  for the boundaries of the NEE phase) coincide with the ones obtained in Ref.~\cite{monthus-LRP} within the Wigner-Weisskopf approximation for $1 < \mu < 2$.
In the present paper we complete the study of the model by providing new detailed results on several observables related to the level statistics, the statistics of the wave-functions's amplitudes, and the statistics of the LDoS in all the regions of the phase diagram.
 

The rest of the paper is organized as follows: In the next section we define the model; In Sec.~\ref{sec:PD} we put forward a novel 
physically transparent argument that allows one to determine the multifractal structure of the mini-bands and the anomalous dimensions of the eigenstates in the NEE phase, as well as the phase diagram of the L-RP ensemble; 
In Sec.~\ref{sec:estimation} we discuss simple ``rules of thumb'' criteria for localization and ergodicity of dense random matrix with uncorrelated entries recently formulated in Refs.~\cite{bogomolny,nosov,kravtsov1,khay}; 
In Sec.~\ref{sec:cavity} we investigate the statistics of the local resolvent by means of the cavity approach which fully supports the results presented in the previous sections; In Sec.~\ref{sec:D} we compare the analytical predictions for the fractal exponents with extensive numerical simulations; 
In Secs.~\ref{sec:level} and~\ref{sec:K2} we investigate numerically the behavior of the level statistics and of the spectral correlation functions, showing that they are in full agreement with the theoretical analysis; 
Finally, in Sec.~\ref{sec:conclusions} we present some concluding remarks and perspectives for future investigations.
In the Appendix sections we present some supplementary information that complement the results discussed in the main text, as well as some technical aspects. 

\section{The model} \label{sec:model}

We consider a natural modification of the RP  ensemble~\cite{monthus-LRP} where the independent and identically distributed off-diagonal elements are  taken from  a L\'evy distribution with power-law tails~\cite{benarous,JP,burda,metz,noi,monthus,auff,borde,edges,cauchy}.
The Hamiltonian of the L-RP model is a sum of two independent $N \times N$ matrices:
\begin{equation} \label{eq:H}
{\cal H} = {\cal A} + \kappa {\cal L}_{\mu,\gamma} \, ,
\end{equation}
where ${\cal A}_{ij} = \epsilon_i \delta_{ij}$ is a diagonal matrix with i.i.d. random entries taken from a given distribution of width $W$ (our results are independent of its specific form~\cite{footnote3}), and ${\cal L}_{\mu,\gamma}$ is a L\'evy matrix with i.i.d broadly distributed elements with a power-law tail of exponent $1 + \mu$ and typical value of the order $N^{-\gamma/\mu}$. 
$\kappa$ is a constant of $O(1)$.
For concreteness one can take a student distribution which reads:
\begin{equation} \label{eq:Llevy}
P_{\mu,\gamma} ({\cal L}_{ij}) = \frac{\mu}{2 N^\gamma {\cal L}_{ij}^{1+\mu}} \, \theta (|{\cal L}_{ij}| > N^{-\gamma/\mu}) \, .
\end{equation}
The largest elements of each row or column of ${\cal L}_{\mu,\gamma}$ are of order $N^{(1-\gamma)/\mu}$. Hence, they are much smaller than $W$ for any $\gamma > 1$ (while the largest element of  the whole $N \times N$ matrix 
is of order  $N^{(2-\gamma)/\mu}$, which is much smaller than $W$ only for $\gamma > 2$). 
The average  
DoS of ${\cal H}$ is thus given by the DoS of ${\cal A}$, $\rho(E) \simeq p(E)$, for $\gamma>1$ in the thermodynamic limit (except a vanishing fraction of eigenvalues of energy of order $N^{(2-\gamma)/\mu}$ in the Lifshitz tails of the spectrum). 
For $\gamma<1$, it is instead given by the DoS 
of L\'evy matrices (which can be computed exactly~\cite{benarous}) but with eigenvalues proportional to $N^{(1 - \gamma)/\mu}$.
The standard RP model is recovered for $\mu = 2^+$~\cite{kravtsov, facoetti}, when the variance of  the ${\cal L}_{ij}$'s is finite and the off-diagonal matrix belongs to the GOE ensemble, with the typical value scaling as $[{\cal L}_{ij}]_{\rm typ} \sim N^{-\gamma/2}$. For $\mu>2$ the average DoS is thus given by $p(E)$ for $\gamma>\mu/2$ and by the semicircle law for $\gamma<\mu/2$, with all eigenvalues rescaled by $N^{(\mu-2 \gamma)/(2 \mu)}$.
In summary, in the bulk of the spectrum  the mean level spacing $\Delta$  is:
\begin{equation} \label{eq:mls}
\Delta  \simeq	\left \{ 
\begin{array}{ll}
	W N^{-1} & \textrm{for~} \mu<2 \textrm{~and~} \gamma>1 \, , \\
	\kappa N^{(1 - \gamma - \mu)/\mu} & \textrm{for~} \mu<2 \textrm{~and~} \gamma< 1 \, ,\\
	W N^{-1} & \textrm{for~} \mu>2 \textrm{~and~} \gamma>\mu/2 \, , \\
		\kappa N^{-(2 \gamma + \mu)/(2 \mu)} & \textrm{for~} \mu>2 \textrm{~and~} \gamma< \mu/2 \, ,\\
	\end{array}
	\right. 
	\end{equation}

Since L\'evy matrices  play a central role in the L-RP model, let us recall the main results. L\'evy matrices (corresponding to $W=0$, $\mu<2$, and $\gamma=1$) have been intensively investigated in the last few years both from the mathematical and the physical sides~\cite{benarous,JP,burda,metz,noi,monthus,auff,borde,edges,cauchy,lopatto1,lopatto2}, since they represent a very broad universality class, with different and somehow unexpected properties compared to the Gaussian case. Their phase diagram turns out to be quite rich~\cite{JP,metz,noi}: For $\mu>1$ all eigenvalues in the bulk are fully delocalized and the level statistics is described by the GOE ensemble on the scale of the mean level spacing. There is however a small sub-extensive fraction
of localized eigenvectors corresponding to the $N^{3/(2+\mu)}$ largest eigenvalues in the tails of the spectrum~\cite{monthus}. For $\mu<1$, instead, a mobility edge appears at finite energy, separating extended eigenstates of energy $E < E_{\rm loc} (\mu)$ from localized eigenstates of energy $E > E_{\rm loc} (\mu)$. The statistics of neighboring levels is described by the GOE ensemble for $E < E_{\rm loc} (\mu)$  and by Poisson statistics for $E > E_{\rm loc} (\mu)$. The localization transition taking place at $E_{\rm loc}$ shares all the properties of AL in the tight-binding Anderson model on the Bethe lattice~\cite{abou,susy,Bethe,tikhonov}.
As shown in Ref.~\cite{noi}, the mobility edge can be computed analytically. $E_{\rm loc} (\mu)$ does not depend on $N$ in the thermodynamic limit for the natural scaling of the off-diagonal elements $\gamma=1$ and tend to $0$ for $\mu \to 0$ and diverges for $\mu \to 1^-$.  With the scaling of~(\ref{eq:H}) the mobility edge found for $0< \mu < 1$ thus moves to energies of the order $N^{(1 - \gamma)/\mu}$.

\section{Phase diagram of the L-RP ensemble} \label{sec:PD}

As first shown in Ref.~\cite{kravtsov}, and later further discussed in Refs.~\cite{warzel,facoetti,bogomolny,bera,pino}, the RP random matrix model with diagonal disorder of width $W$ and off-diagonal i.i.d. Gaussian elements of variance $N^{-\gamma}$ has three phases: fully ergodic for $\gamma<1$, 
NEE for $1<\gamma<2$, and fully localized for $\gamma >2$, and two transitions between them at $\gamma_{\rm ergo} = 1$ and $\gamma_{\rm AL} = 2$. The same kind of phases and transitions between them  are expected for the L-RP model~\cite{monthus-LRP}.

\begin{figure}
\includegraphics[width=0.49\textwidth]{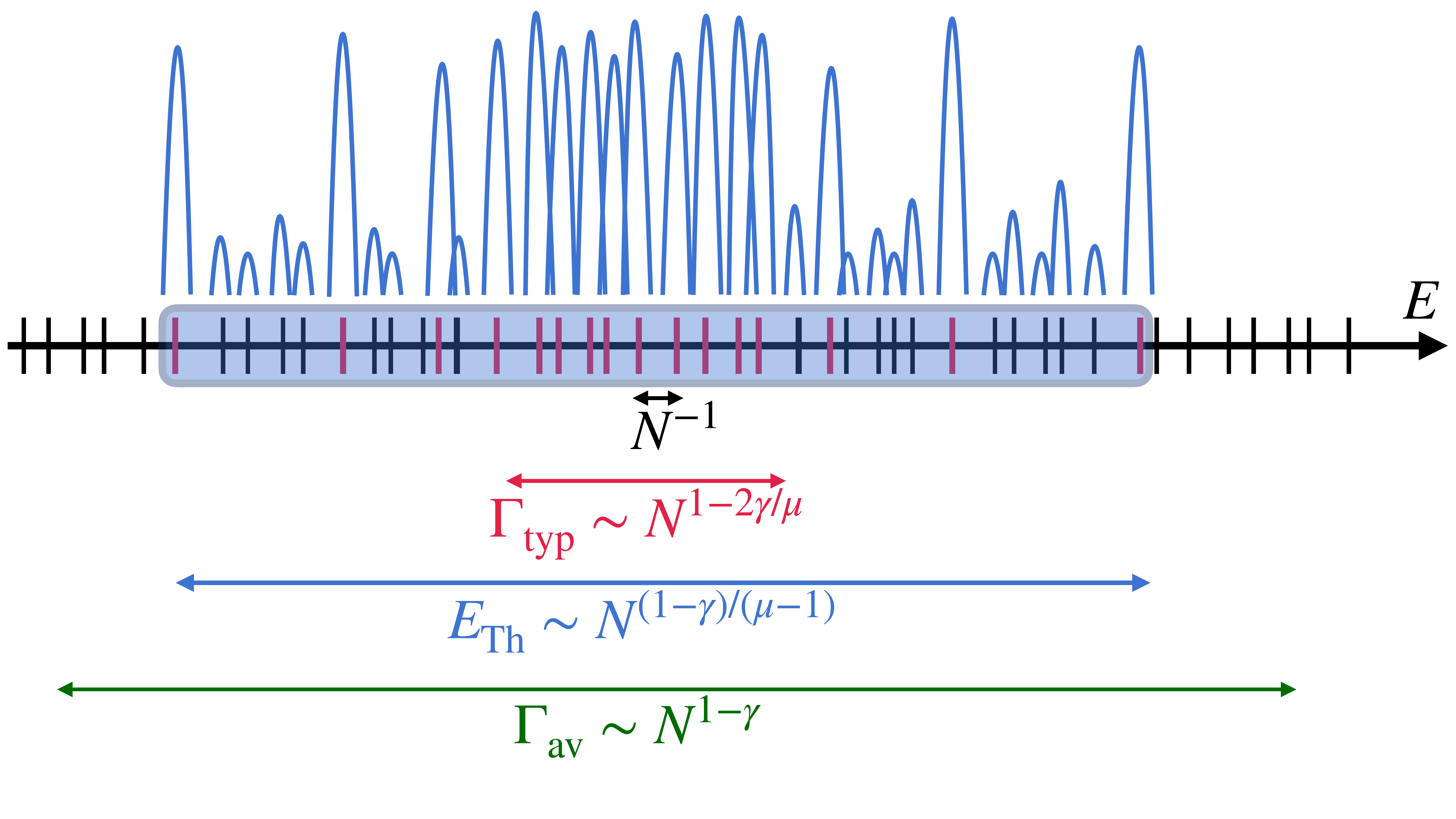}
	
	\vspace{-0.6cm}
	\caption{\label{fig:thouless} Pictorial representation of the fractal structure of the mini-bands and of a wave-function belonging to it for $1<\mu<2$ and $1<\gamma<\mu$. Consecutive levels are in resonance due to typical off-diagonal matrix elements. However the mini-bands extend to a much larger energy scale due to rare resonances produced 
		by anomalously large matrix elements. The energy spreading $E_{\rm Th}$ of the mini-bands is determined self-consistently by asking that the maximum hybridization gap ${\cal L}_{ij}$ of 
		$E_{\rm Th}/N^{-1}$ adjacent levels has the same scaling of the width of the mini-bands itself, Eq.~(\ref{eq:self-consistent}), and is much smaller than the average effective bandwidth $\Gamma_{\rm av}$ found from standard perturbation theory.}
\end{figure}

In order to proceed further, let us first recall that the ``smoking gun'' evidence~\cite{kravtsov,facoetti,bogomolny}  of the NEE  phase is the presence of the mini-bands in the local spectrum: Eigenstates occupy a sub-extensive fraction of the total volume and spread over $N^{D_1}$ consecutive energy levels which are hybridized by the off-diagonal perturbation, while wave-functions belonging to different support sets do not overlap in the thermodynamic limit.
In the Gaussian RP model the width of the mini-bands, 
called the Thouless energy, is given by $E_{\rm Th} \propto \Gamma_{\rm av} = 2 \pi \rho N \langle | {\cal H}_{ij} |^2 \rangle$, 
which is, according to the Fermi golden rule, the average escape rate of a particle at a given site $i$ ($\rho$ is the average DoS of ${\cal H}$, see Sec.~\ref{sec:estimation} for more details). In the NEE phase ($1 < \gamma <2$) one has that $E_{\rm Th} \sim N^{1 - \gamma}$, with $\Delta \ll E_{\rm Th} \ll W$. On the other hand the Thouless energy must be also equal to the number of sites of a support set occupied by an eigenstate, $N^{D_1}$, times the average distance between  consecutive levels, $\Delta \sim N^{-1}$, implying that $D_1 = 2 - \gamma$.
As anticipated in the introduction, the mini-bands in the Gaussian RP model are {\it fractal} (and not {\it multifractal}) and the anomalous dimensions are not-degenerate ($D_q = D$ $\forall q>1/2$). 

Below we illustrate a very simple,  general,  and physically transparent   argument that yields $E_{\rm Th}$ and $D_1$ when hybridization occurs on an energy scale much larger than  $\Gamma_{\rm typ}$ due to anomalously large  matrix elements in the tail of the distribution  (see Fig.~\ref{fig:thouless} for a pictorial illustration of this argument).
Besides the structure of the mini-bands in the NEE phase and the  anomalous dimensions which characterize the statistics of the amplitudes of the multifractal wave-functions, this argument also yields the phase diagram of the L-RP ensemble. 
These predictions will be then confirmed in the next sections both analytically, by a thorough analysis of the cavity equations, and numerically, by means of extensive exact  diagonalizations.

	Let us focus on the case $\mu<2$ and $\gamma>1$ and let us assume that  in the NEE phase the mini-bands contain $N^{D_1}$ energy levels, and thus extend up to an energy scale of $E_{\rm Th}=N^{D_1} \times N^{-1}$. 
	Let us consider a site $i$ belonging to a given mini-band. Hybridization of $i$ with the  $N^{D_1}$ levels  $j$ of the support set is  only possible if the maximum of the off-diagonal matrix elements ${\cal H}_{ij}$ among those levels,  which scales as $N^{D_1/\mu - \gamma/\mu}$, is of the same order of the width of the mini-band itself: 
	\begin{equation} \label{eq:self-consistent}
		E_{\rm Th} \sim N^{D_1 - 1} \sim  {\rm max}_{j=1,\ldots,N^{D_1}}  \{ {\cal H}_{ij} \} \sim N^{D_1/\mu - \gamma/\mu} \, .
	\end{equation}
	Hence in the  NEE phase one must have $D_1 = (\mu-\gamma)/(\mu - 1)$ and $E_{\rm Th} \sim N^{(1 - \gamma )/ (\mu - 1)}$. (The expression found for $E_{\rm Th}$ is in fact in agreement with the one of Ref.~\cite{monthus-LRP}, although it is has been obtained with a different approach. Conversely the approximations of Ref.~\cite{monthus-LRP} do not lead to the correct result for the fractal dimensions, which were predicted to be equal to zero for all $q>\mu/2$.) 
	
	Anderson localization  occurs when the mini-bands' width formally becomes smaller than the mean level spacing. At this point, which corresponds to $E_{\rm Th} \sim N^{-1}$, i.e. $\gamma_{\rm AL} = \mu$ for $\mu>1$~\cite{monthus-LRP}, the levels of $\mathcal A$ are almost unaffected by the Levy perturbation. 
	Conversely, ergodicity is restored when the Thouless energy becomes of the order  of the total spectral bandwidth, $E_{\rm Th} \sim W$, i.e., $\gamma_{\rm ergo} =1$~\cite{monthus-LRP}.
	
\begin{figure}
		\begin{center}
			\includegraphics[width=0.49\textwidth]{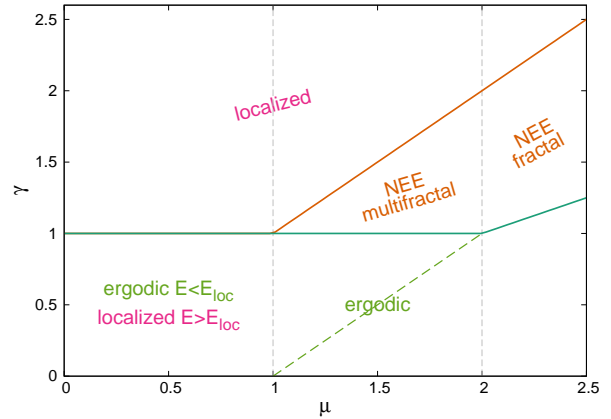}
			\vspace{-1cm}
			\caption{\label{fig:PD} Phase diagram of the L-RP ensemble. 
				The transition lines to the AL phase 
				$\gamma_{\rm AL}$ 
				and to the ergodic phase 
				$\gamma_{\rm ergo}$ are given in Eq.~(\ref{eq:gammaergoAL}). 
				For $\mu>2$ 
				the NEE phase becomes fractal (and not multifractal). As shown in Ref.~\cite{noi}, for $\mu<1$ and $\gamma<1$ a mobility edge separates a delocalized phase for $E<E_{\rm loc}$ from a Anderson localized phase for $E>E_{\rm loc}$ with $E_{\rm loc} \sim N^{(1-\gamma)/\mu}$ with the  scaling of Eq.~(\ref{eq:Llevy}).
				The green dashed line separates a ``weakly'' ergodic regime, $(\mu<1 ,E<E_{\rm loc})$ and $(1<\mu<2, \gamma>1 + \mu)$, similar to the metallic phase of the Anderson model (see Sec. IV and \cite{khay} for its definition), from a fully ergodic one $(1<\mu<2,\gamma>1 + \mu)$ and $(\mu>2 ,\gamma<\mu/2)$, where the  orthogonal symmetry fully establishes.
			} 
		\end{center}
	\end{figure}

	For $\mu \to 2$ one recovers the RP result, $E_{\rm Th} \sim N^{1 - \gamma}$, and the Thouless energy becomes equal to the typical effective bandwidth.
	For $\mu>2$ one thus has that $E_{\rm Th} \sim \Gamma_{\rm typ} \sim N^{1 - 2 \gamma/ \mu}$ and $D_1 = 2 - 2 \gamma / \mu$.
	At the AL transition $E_{\rm Th} \sim \Delta$ [given in Eq.~(\ref{eq:mls})], i.e. $\gamma_{\rm AL} = \mu$, while ergodicity is fully restored when $E_{\rm Th}\sim W$, i.e. $\gamma_{\rm ergo} = \mu/2$.
	The resulting phase diagram of the L-RP ensemble is reported in Fig.~\ref{fig:PD}, and the transition lines between the different phases are:
	\begin{equation} \label{eq:gammaergoAL}
		\gamma_{\rm ergo} = 
		\left \{
		\begin{array}{ll}
			1 & \textrm{for~} \mu \le 2 \, , \\
			\mu/2 &  \textrm{for~} \mu > 2 \, , 
		\end{array}
		\right .
		\qquad \gamma_{\rm AL} = 
		\left \{
		\begin{array}{ll}
			1 & \textrm{for~} \mu \le 1 \, , \\
			\mu &  \textrm{for~} \mu > 1 \, . 
		\end{array}
		\right .
	\end{equation}
	This analysis predicts  the existence of a tricritical point  (similarly to the LN-RP ensemble~\cite{kravtsov1,khay}) for $\mu = 1$ (i.e., Cauchy distributed off-diagonal elements~\cite{cauchy}) and $\gamma=1$. We will come back to the peculiar properties of such tricritical point in the next sections.
	
In the larger $N$ limit the spectral  dimensions $D_1$ is thus given by:
\begin{equation} \label{eq:D1}
	D_1 (\gamma) = \left \{
	\begin{array}{ll}
		1 & \textrm{for~} \gamma \le \gamma_{\rm ergo} \, \\
		\frac{\mu - \gamma}{\mu - 1} & \textrm{for~}  \mu<2 \textrm{~and~}  \gamma_{\rm ergo} <\gamma  \le \gamma_{\rm AL} \, , \\
		2 - \frac{2 \gamma}{ \mu} &  \textrm{for~}  \mu>2 \textrm{~and~}  \gamma_{\rm ergo}  <\gamma  \le \gamma_{\rm AL}  \, , \\
		0 & \textrm{for~} \gamma > \gamma_{\rm AL}  \, ,
	\end{array}\right .
\end{equation}
implying that $D_1$ is continuous for $\mu > 1$ both at the transition to the NEE phase ($D_1 \to 1$ for $\gamma \to \gamma_{\rm ergo}^+$) and at the Anderson transition ($D _1\to 0$ for $\gamma \to \gamma_{\rm AL}^-$).
For $\mu \le 1$, instead $D_1$ is expected to display a discontinuous jump from $D_1=1$ for $\gamma =1$ to $D_1 =0$ for $\gamma=1^+$~\cite{footnote1}.

	The argument illustrated above also suggests that 
	higher order moments  of the wave-functions' amplitudes exhibit a different scaling with $N$, and hence that the fractal dimensions are not degenerate.
		In particular, 
		the anomalous dimension $D_{\infty}$, associated with the scaling of the number of sites where the  amplitudes take the largest values, should be dominated by the compact part of the support set of the eigenstates.
		In fact most of the sites where the wave-functions' amplitudes are large falls within the typical bandwidth and only few of them (i.e., a sub-extensive fraction of the $N^{D_1}$ sites of the support set) are outside it (see Fig.~\ref{fig:thouless}). On most of the sites of the mini-band at energy separation larger than $\Gamma_{\rm typ}$ the amplitudes are typically smaller since these sites are only hybridized at higher orders in perturbation theory.
		One thus has:  
		\begin{equation} \label{eq:Dinf}
			D_\infty (\gamma) = \left \{
			\begin{array}{ll}
				1 & \textrm{for~} \gamma \le \gamma_{\rm ergo} \, \\
				2 - \frac{2 \gamma}{\mu } & \textrm{for~} \gamma_{\rm ergo} <\gamma  \le \gamma_{\rm AL}   \, , \\
				0 & \textrm{for~} \gamma > \gamma_{\rm AL}  \, .
			\end{array}\right .
		\end{equation}
		This implies that $D_\infty$  should  display a finite jump at the ergodic transition for $\mu<2$. The amplitude of the jump is $1 - 2 / \mu$ and goes to one for $\mu \to 1$ and to zero in the RP limit, $\mu \to 2$.
		The difference between $D_1$ and $D_\infty$ confirms that the L-RP ensemble displays {\it multi}fractal behavior, contrary to the Gaussian RP model. 
In the latter, 
one has that $\langle | {\cal H}_{ij} |^2 \rangle = [ {\cal H}_{ij}]^2_{\rm typ}$, implying that the {\it average} effective spectral bandwidth  coincides with the {\it typical} one. 
This is of course not the case for the L-RP ensemble, since the matrix elements are broadly distributed and $\Gamma_{\rm av} \neq \Gamma_{\rm typ}$ for $\mu<2$.
		
		Note that since neighboring energy levels are always hybridized by the off-diagonal terms, the level statistics on the scale of the mean level spacings is expected to be described by the GOE ensemble in the whole NEE phase, as for the Gaussian RP model.
		In Secs.~\ref{sec:cavity}-\ref{sec:K2} we will present a thorough analytical and numerical investigation of the L-RP model that  fully confirms the predictions of 
		Eqs.~(\ref{eq:gammaergoAL}),~(\ref{eq:D1}), and ~(\ref{eq:Dinf}), while
		in the next section we show that the phase diagram of Fig.~\ref{fig:PD} is  in agreement with the simple ``rules of thumb'' criteria for localization and ergodicity of dense random matrices with uncorrelated entries recently formulated in Refs.~\cite{bogomolny,nosov,kravtsov1,khay}.
		In App.~\ref{sec:first} we show that in the AL phase, where  perturbation theory converges absolutely, one can determine the whole spectrum of fractal dimensions  exactly, Eq.~(\ref{eq:Dloc}).
		
		Note however that the argument presented above do not take into account the spectral properties of the off-diagonal L\'evy perturbation. In fact, as explained in Sec.~\ref{sec:model}, 
		for $\mu<1$ a mobility edge appears in the spectrum of L\'evy matrices, 
		separating fully extended eigenstates of energy $E < E_{\rm loc} (\mu)$ from AL eigenstates of energy $E > E_{\rm loc} (\mu)$~\cite{metz,noi}.  
		For $\gamma=1$ the mobility edge $E_{\rm loc} (\mu)$ is finite and do not depend on $N$ in the thermodynamic limit ($E_{\rm loc} (\mu)$  tends to $0$ for $\mu \to 0$ and diverges for $\mu \to 1^-$~\cite{noi}). 
		With the scaling of~(\ref{eq:H}) the mobility edge  thus moves to energies of the order $N^{(1 - \gamma)/\mu}$.
		Since 
		the rare large off-diagonal elements which are responsible of the hybridization of the energy levels are in fact associated to strongly localized eigenfunctions, 
		one 
		expects that for $\mu<1$ and $\gamma<1$ the system is delocalized at low energy and localized at high energy, with a mobility edge scaling as $N^{(1 - \gamma)/\mu}$. We will  study this region of the phase diagram in App.~\ref{sec:mu1}.
		
		The other aspect that the argument might not take into account is that, as recently shown in~\cite{khay}, the multifractal states might be fragile against hybridization and the NEE phase could be in fact squeezed due to this effect. We will investigate this possibility in App.~\ref{sec:stability}, showing that for the L-RP ensemble such instability does not take place.
		
		We would like to stress the fact that  the argument presented in this section is very general and physically transparent and can be in principle extended to analyze the multifractal states in other systems. The same kind of ideas and reasoning might be reformulated and adapted to situations in which the matrix elements are  correlated~\cite{nosov} and/or depend on the matrix indices and on the energy separation, 
		as in more realistic interacting models~\cite{qrem3,tarzia,war}. As an illustration, in App.~\ref{app:LNRP} we show  that applying these ideas to the LN-RP ensemble of Refs.~\cite{kravtsov1,khay} allows one to obtain the phase diagram and the anomalous dimensions of the model in a few-lines 
		calculation.

\section{Simple ``rules of thumb'' criteria for localization and ergodicity} \label{sec:estimation}

 In this section we apply  the ``rules of thumb'' criteria for localization and ergodicity of dense random matrix with uncorrelated entries recently formulated in Refs.~\cite{bogomolny,nosov,kravtsov1,khay}, showing that they  yield  an estimation of the phase diagram of the L-RP ensemble and of the transition lines between the different phases which are in full agreement with Fig.~\ref{fig:PD} and Eq.~(\ref{eq:gammaergoAL}). 

The first criterion~\cite{bogomolny,nosov,kravtsov1,khay}  states that Anderson localization occurs when the sum:
\begin{equation} \label{eq:AL}
\lim_{N \to \infty} \sum_{j=1}^N \langle | {\cal H}_{ij} | \rangle < \infty\, .
\end{equation}
The physical interpretation of this condition is that if the number of sites $j$  in resonance with a given site $i$ is finite in the thermodynamic limit then the system is localized.

The second criterion~\cite{bogomolny,nosov,kravtsov1,khay}  is a sufficient condition of  ergodicity. It states that if the sum:
\begin{equation} \label{eq:FGR}
\lim_{N \to \infty} \sum_{j=1}^N \langle | {\cal H}_{ij} |^2 \rangle \to \infty 
\end{equation}
 the system is ergodic.
Its physical interpretation is obtained by recalling that, according to the Fermi Golden Rule, the spreading amplitude
\begin{equation} \label{eq:mb}
	\Gamma_{i} 
	\approx 2 \pi \rho \sum_j 
	| {\cal H}_{ij} |^2 
\end{equation}
quantifies the escape rate of a particle created at a given site $i$, where $\rho$ is the average DoS of ${\cal H}$. 
For $\mu<2$ and $\gamma>1$ one can neglect contribution of off-diagonal matrix elements to the density of states and $\rho (E) \simeq p(E)$, 
and  the total spectral bandwidth is limited by $W$. 
The condition~(\ref{eq:FGR}) thus states that when the average spreading width $\Gamma_{\rm av}$ is much larger than the spread of energy levels $W \sim O(1)$ due to disorder, then the system is in the ergodic phase since starting from a given site the wave-packet spreads to any other given site in times of order one. 
In other words, the fulfillment of this condition implies that there are no mini-bands in the local spectrum. 
[For $\gamma<1$ and $\mu<2$ or $\gamma<\mu/2$ and $\mu>2$ instead, $\Gamma_{\rm av} \gg W$ and 
the total spectral bandwidth $B = N \Delta$ 
is given by the off-diagonal matrix elements~\cite{noi}, see Eq.~(\ref{eq:mls}).]
 
 The  NEE phase is thus realized if:
 \[
 \lim_{N \to \infty} \sum_{j=1}^N \langle | {\cal H}_{ij} | \rangle \to \infty \, , \,\,\,\,  \textrm{and} \,\,\,\,\, \lim_{N \to \infty} \sum_{j=1}^N \langle | {\cal H}_{ij} |^2 \rangle \to 0 \, . 
 \]
For L\'evy distributed off-diagonal elements the second moment of $|{\cal H}_{ij}|$ diverges for any $\mu <2$ and the first moment diverges for any $\mu<1$.
However, 
the averages appearing in Eqs.~(\ref{eq:AL}) and~(\ref{eq:FGR})  should be done with the distribution truncated at the total spectral bandwdth $B = N \Delta$ [where $\Delta$ is given in Eq.~(\ref{eq:mls})].
The reason for that is that rare large matrix elements $|{\cal H}_{ij}| \gg B$ split the resonance pair of levels so much that they are pushed at the Lifshitz tail of the spectrum and do not affect statistics of states in its bulk~\cite{kravtsov1,khay}.
We thus have that: 
\[
 \langle |{\cal H}_{ij} |^q \rangle_B \sim 
	\left \{
	\begin{array}{ll}
		\frac{\mu \kappa^q}{\mu - q} N^{- \gamma q /\mu} & \textrm{for~} q < \mu  \, , \\ 
		\gamma \kappa^q N^{- \gamma} \log N & \textrm{for~} q = \mu  \, , \\ 
		\frac{\mu \kappa^q B^{q-\mu}}{q -\mu} N^{- \gamma} & \textrm{for~} q > \mu \, .
	\end{array}
	\right .
	\]
Applying the criteria~(\ref{eq:AL}) and~(\ref{eq:FGR})  one  thus immediately recovers the phase diagram of Fig.~\ref{fig:PD} and the transition lines given in Eq.~(\ref{eq:gammaergoAL}).
Note that at the tricritical point  ($\mu = 1$, $\gamma=1$)  one has that
$N \langle |{\cal H}_{ij} | \rangle_W  \sim \log N\to \infty$ and $N \langle {\cal H}_{ij}^2 \rangle_W \sim W$. Hence the tricritical point is in the delocalized phase and should be characterized by a very weak ergodicity (wave-functions occupy a finite fraction of the total Hilbert space). On the  line of critical points at $\gamma=1$ for $0<\mu<1$ 
both $N \langle |{\cal H}_{ij} | \rangle_W $ and $N \langle {\cal H}_{ij}^2 \rangle_W$ are finite and $\mu$-dependent with the ratio $\langle |{\cal H}_{ij} | \rangle_W/\langle {\cal H}_{ij}^2 \rangle_W \to \infty$ for $\mu \to 1$.
Similarly, on the line of critical points separating the ergodic regime from the NEE one at $\gamma=1$ and $1<\mu<2$ one has that
$N \langle |{\cal H}_{ij} | \rangle_W  \sim N^{1-1/\mu} \to \infty$ while $N \langle {\cal H}_{ij}^2 \rangle_W \sim W^{2-\mu}/(2-\mu)$. Hence such critical line is in the ergodic phase and should be characterized by $D_q=1$. 

Although the criteria~(\ref{eq:AL}) and~(\ref{eq:FGR}) are originally based on first and second-order perturbation theory for the eigenvectors and the eigenvalues, they give the correct results for the transition lines of the L-RP ensemble. Nevertheless, 
as discussed above, 
 differently from the Gaussian RP counterpart, $N \langle {\cal H}_{ij}^2 \rangle_B$ does not necessarily 
coincide with  $N \langle |{\cal H}_{ij} | \rangle_B^2$ due to the heavy-tails of the distribution of the transition amplitudes.  
This property has several important consequences:\\
1) The first implication 
is that  the energy band $\Gamma_{\rm av}$ obtained from the FGR, which corresponds to the average spreading of the energy levels  due to the off-diagonal perturbation, can be much larger than the mean level spacing in the AL phase. 
In fact for $\mu<2$ and $\gamma_{\rm AL} < \gamma <2$ one has that $\Gamma_{\rm av} \gg N^{-1}$.
In particular, on the transition line $\gamma=\mu$ for $1<\mu<2$ one finds that $N \langle {\cal H}_{ij}^2 \rangle_W \sim N^{1 - \mu}$, while  on the transition line $\gamma=1$ for $0<\mu<1$ one finds that $N \langle {\cal H}_{ij}^2 \rangle_W \sim O(1) $. This is a clear manifestation of the failure of the perturbative expansion (see App.~\ref{sec:first} for more details): Although the energy levels are scrambled by the matrix ${\mathcal L}_{\mu,\gamma}$ by a huge amount compared to $\Delta$, 
the system is nevertheless localized due to the fact that different eigenstates do not overlap and cross each other without interacting.\\
2) A second consequence, already anticipated above and further discussed in Secs.~\ref{sec:cavity} and~\ref{sec:D}, 
is the fact that the typical escape rate $\Gamma_{\rm typ} \approx 2 \pi \rho N [{\cal H}_{ij}]_{\rm typ}^2 \sim N^{1 - 2 \gamma / \mu}$ does not 
coincide with the average one $\Gamma_{\rm av} \sim N^{1 - \gamma}$ for $\mu<2$. This means that 
the {\it typical} energy band hybridized by the off-diagonal perturbation is much smaller than the {\it average} spreading width, 
which is a clear signature of the multifractality of the mini-bands  in the NEE phase 
(see Fig.~\ref{fig:thouless}).\\
3) Finally,  the fact that $\Gamma_{\rm typ} \neq \Gamma_{\rm av}$ have also some implications on the properties of  the ergodic phases, and have led the authors of Ref.~\cite{khay} to put forward an extra  sufficient criterion for ``full ergodicity'' which states that it is realized if
\begin{equation} \label{eq:fullergo}
\lim_{N \to \infty} \frac{\left ( N [ {\cal H}_{ij} ]^2_{\rm typ} \right)^2}{ N \langle | {\cal H}_{ij} |^2 \rangle_B} \to \infty \, .
\end{equation}
If this condition is not full-filled 
the eigenfunction statistics is not invariant under basis rotation and the Wigner-Dyson statistics only establishes up to a finite energy scale, corresponding to a ``weakly'' ergodic phase in which the typical DoS is smaller than the average DoS. This is, for instance, what happens in the ergodic phase of L\'evy matrices~\cite{noi} or in the metallic phases of the Anderson model in three dimensions~\cite{kramir1,kramir2} and on the Bethe lattice~\cite{Bethe,tikhonov}, in which the Thouless energy is finite but strictly smaller than the total spectral bandwidth and the GOE statistics only establishes up to an energy scale of $O(1)$.
Conversely, if~\eqref{eq:fullergo} is verified, i.e. $\mu > 2$ and $\gamma<\mu/2$ or $1 < \mu  < 2$ and $\gamma < \mu - 1$,  the rotation invariance of the GOE ensemble fully establishes in ergodic phase. 

\section{Cavity equations and local resolvent statistics} \label{sec:cavity}
Using the cavity method (or, equivalently, the block matrix inversion formula), it is possible to derive the equations 
relating the probability distribution of the diagonal elements of the resolvent of matrices of size  $N+1$ to those of size $N$~\cite{JP,metz,noi,facoetti,bogomolny}. In the large $N$ limit these equations become asymptotically exact and read:
\begin{equation} \label{eq:cavity}
\left [ {\cal G}_{ii}^{(N+1)} \right]^{-1} = \epsilon_i - E - i \eta - \sum_{j=1}^N {\cal L}_{ij}^2 {\cal G}_{jj}^{(N)} \, .
\end{equation}
In Ref.~\cite{facoetti} it was shown that the existence of the NEE phase of the standard RP model
can be revealed studying a non-standard scaling limit in
which the small additional imaginary regulator $\eta$ vanishes as $N^{-\delta}$.
At the Thouless energy $E_{\rm Th}$---which is proportional to the typical level spacing, $N^{-1}$, times the number of sites, $N^{D_1}$, over which the eigenvectors are delocalized---the spectral statistics displays a crossover from a behaviour characteristic of standard localized phases to a behaviour similar to the one of standard delocalized phases. 
Thus, inspecting the local resolvent statistics one has a direct access to the non-ergodic properties of the delocalized phase.

In the following we carry out a similar analysis for the L-RP ensemble, focusing on the region $\mu < 2$ of the phase diagram (the case $\mu>2$ is analogous to the one discussed in Refs.~\cite{facoetti,bogomolny}). 
We focus on the imaginary part of the Green's function
and drop the $N$-dependence in Eq.~(\ref{eq:cavity}) since one can assume that in the thermodynamic limit the distribution of ${\cal G}^{(N+1)}$ is the same as the distribution of ${\cal G}^{(N)}$.

The matrix elements ${\cal L}_{ij}$ and the elements of the resolvent are by construction uncorrelated. In the Gaussian RP case~\cite{facoetti,bogomolny} from the central limit theorem one has that the imaginary part of the self energy 
	\[
	S_i (z) = \sum_{j=1}^N {\cal L}_{ij}^2 \, {\rm Im} {\cal G}_{jj} (z)
	\]
	 (with $z=E- i \eta$) is a Gaussian variable whose expectation value is given by $\pi \rho (E) N^{1-\gamma}$, where $\rho(E) \simeq p(E)$ is the average DoS. The scaling of the self energy thus sets the scale of the Thouless energy, $E_{\rm Th} \propto N^{1-\gamma}$, on which the crossover from localized-like behavior (for $\eta \gg N^{1-\gamma}$) to  delocalized-like behavior (for $\eta \ll N^{1-\gamma}$) takes place~\cite{facoetti}.
The situation is however   more involved  in the L-RP case, where the  ${\cal L}_{ij}^2$'s are broadly distributed. Using the generalized central limit theorem one finds that $S_i(z)$  is a L\'evy distributed random variable with power law tails with an exponent $1 + \mu/2$ and typical value given by
\begin{equation} \label{eq:Styp}
\begin{aligned}
[S(z)]_{\rm typ} &= 
N^{\frac{2(1-\gamma)}{\mu}}  \left [ \left \langle \left \vert  {\rm Im} {\cal G} (z) \right \vert^{\frac{\mu}{2}} \right \rangle \right]^{\frac{2}{\mu}} \, .
\end{aligned}
\end{equation}
Hence, the typical value of the self energy is related to the $(\mu/2)$-th moment of the Green's function and must be determined self-consistently, as explained in the following. 
Neglecting the real part of the resolvent (by using the same arguments as the ones given below one can in fact show that the real parts only give a subleading contribution) we get:
\begin{equation} \label{eq:ImG}
{\rm Im} {\cal G}_{ii} (z) \simeq \frac{\eta + S_i(z)}{[E - \epsilon_i]^2 + [\eta + S_i(z)]^2} \, .
\end{equation}
Let us imagine a situation in which $\eta \gg S_i$: 
\[
{\rm Im} {\cal G}_{ii}  \simeq \left \{
\begin{array}{ll}
\frac{\eta}{(E - \epsilon_i)^2} & \textrm{if~} | E - \epsilon_i | \gg \eta \, , \\
\frac{1}{\eta} & \textrm{if~} | E - \epsilon_i | \ll \eta \, .
\end{array}
\right .
\]
For $N$ large and $\eta$ small  and fixed (but much larger than the $S_i$'s) one then recovers the standard localized behavior:
\begin{equation} \label{eq:ploc}
P({\rm Im} {\cal G}) \propto p(E) \frac{\sqrt{\eta}}{({\rm Im} {\cal G})^{3/2}} \, ,
\end{equation}
with a cutoff at ${\rm Im} {\cal G} = 1/\eta$. For $\mu>1$ we then have
\begin{equation} \label{eq:ImGmutyp}
\left \langle \left \vert  {\rm Im} {\cal G} \right \vert^{\frac{\mu}{2}} \right \rangle \propto \frac{\eta^{1 - \frac{\mu}{2}}}{\mu - 1} \, .
\end{equation}
In fact the $\mu/2$-th moment of ${\rm Im} {\cal G}$ is dominated by the upper cutoff of the distribution for $\mu>1$,  and is given by $(1/\eta)^{\mu/2}$ times the probability to find a resonance such that $| E - \epsilon_i | < \eta$, which is proportional to $\eta$. 
For $\mu<1$, instead, $\langle \vert  {\rm Im} {\cal G}  \vert^{\frac{\mu}{2}} \rangle \propto \eta^{\mu/2}$, and $S_i$ is always negligible with respect to $\eta$ in the denominator of~(\ref{eq:ImG}) as soon as $\gamma>1$. The system is then in the AL phase for $\mu<1$ and $\gamma>1$, in agreement with the results given in the previous section and illustrated in the phase diagram of Fig.~\ref{fig:PD}. Hence in the following we will focus on the range $1<\mu <2 $ only.

\begin{figure}
	\includegraphics[width=0.49\textwidth]{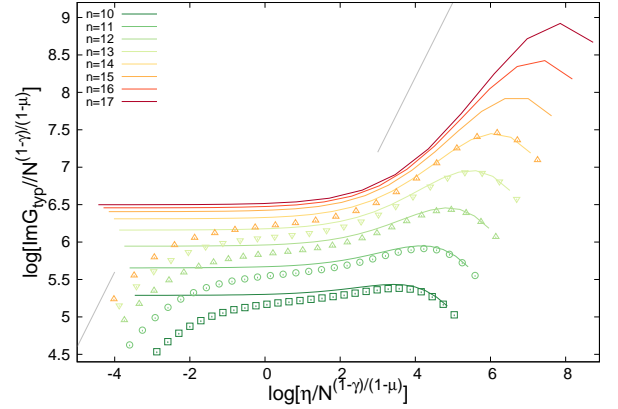}
		\vspace{-1cm}
	\caption{\label{fig:th-cavity} ${\rm Im} {\cal G}_{\rm typ}$ as a function of the imaginary regulator $\eta$ for several system sizes $N=2^n$ (with $n$ from $10$ to $17$) within the NEE phase ($\mu=1.75$ and $\gamma=1.6$), obtained by solving numerically the cavity equation~(\ref{eq:cavity}). Both axis are rescaled by the Thouless energy $N^{(1 - \gamma)/(\mu-1)}$. We also show the results of exact diagonalizations up to $n=14$ (empty symbols) which are in good agreement with the cavity calculation except in the regime $\eta \ll 1/N$, as expected, due to the fact that the spectral statistics of finite size matrices always appears as localized if $\eta$ is smaller than the mean level spacing. The gray lines corresponds to the localized behavior ${\rm Im} {\cal G}_{\rm typ} \propto \eta$ found for $\eta \ll N^{-1}$ and for $\eta \gg E_{\rm Th}$. Similar results are found within the whole NEE phase $1 < \mu < 2$ and $1 < \gamma < \mu$.}
\end{figure}

Let us consider now  the sites where $S_i \gg \eta$, where 
\[
{\rm Im} {\cal G}_{ii}  \simeq \left \{
\begin{array}{ll}
\frac{S_i}{(E - \epsilon_i)^2} & \textrm{if~} | E - \epsilon_i | \gg S_i \, , \\
\frac{1}{S_i} & \textrm{if~} | E - \epsilon_i | \ll S_i \, .
\end{array}
\right .
\]
If $\eta$ is smaller than the typical value of $S_i$, then ${\rm Im} {\cal G}$ becomes independent of $\eta$ on all sites, and its probability distribution is given by Eq.~(\ref{eq:ploc}) with $\eta$ replaced by $S_i$. 
The $\mu/2$-th moment of ${\rm Im} {\cal G}$ must then be determined self-consistently from Eqs.~(\ref{eq:Styp}) and~(\ref{eq:ImGmutyp}), yielding:
\[
\begin{aligned}
\left \langle \left \vert  {\rm Im} {\cal G} \right \vert^{\frac{\mu}{2}} \right \rangle & \propto N^{- \frac{\gamma-1}{\mu - 1} \left( 1 - \frac{\mu}{2} \right) } \, , \\
[ S(z) ]_{\rm typ} & \propto N^{- \frac{\gamma - 1}{\mu - 1}} \, .
\end{aligned}
\]
If instead $\eta$ is larger than $[S(z)]_{\rm typ}$, on most of the sites the regulator dominates over $S_i$. 
We thus have that:
\begin{equation} \label{eq:Th}
{\rm Im} {\cal G}_{\rm typ} \propto \left \{
\begin{array}{ll}
N^{(1 - \gamma)/(\mu-1)} & {\rm for~} \eta \ll N^{(1 - \gamma)/(\mu-1)}  \, , \\
\eta & {\rm for~} \eta \gg N^{(1 - \gamma)/(\mu-1)}  \, .
\end{array}
\right .
\end{equation}
This behavior is confirmed by Fig.~\ref{fig:th-cavity}, where we plot the typical value of the imaginary part of the Green's function obtained by solving numerically Eqs.~(\ref{eq:cavity}) for several values of the regulator $\eta$ and for several system sizes $N=2^n$ (with $n$ from $10$ to $17$) within the intermediate phase ($\mu=1.75$ and $\gamma=1.6$). 
The figure shows that for $N$ large the curves corresponding to different size  approach a limiting curve when  the $\eta$ and the ${\rm Im} {\cal G}_{\rm typ}$ axis are rescaled by the Thouless energy 
$N^{(1 - \gamma)/(\mu-1)}$, as predicted by Eq.~(\ref{eq:Th}).  In particular the plateau establishing for $\eta \ll E_{\rm Th}$ is clearly visible, although some finite-size effects are still at play even for the largest system size $N=2^{17}$. We also show the results of exact diagonalizations up to $N=2^{14}$ which are in very good agreement with the cavity solution, except in the regime $\eta \ll 1/N$. In fact, when the regulator becomes much smaller than the mean level spacing, finite-size L-RP matrices exhibit again the localized behavior, ${\rm Im} {\cal G}_{\rm typ} \propto N^{D_1} \eta$~\cite{prefactor}, as expected. 

This analysis 
reveals the existence of a crossover energy scale $E_{\rm Th} \simeq [ S(z) ]_{\rm typ}  \propto N^{(1 - \gamma)/(\mu-1)}$  over which ${\rm Im} {\cal G}_{\rm typ}$ has a delocalized-like behavior and is independent of $\eta$, in full agreement with the results given in Sec.~\ref{sec:PD}. The origin of such crossover scale is due to the fact that wave-functions close in energy 
are hybridized by the off-diagonal perturbation and form mini-bands. Within the cavity approach the effective with of the mini-bands is  self-consistently determined by finding the width of the energy interval such that $| E - \epsilon_i | \lesssim S_i$.
 Eq.~(\ref{eq:Th}) also  yields a prediction for the spectral fractal exponent $D_1$. 
 In fact, by definition one has that
 \[
 \rho_{\rm typ} = \frac{e^{\langle \log {\rm Im} {\cal G} \rangle} } { \langle {\rm Im} {\cal G} \rangle} \propto N^{D_1 - 1}\, .
 \]
 Since $\langle {\rm Im} {\cal G} \rangle \sim \pi p(E)$ is of order $1$ in the whole intermediate NEE phase (as well as in the AL phase), from the asymptotic behavior of the Green's functions 
 one immediately finds that in the large $N$ limit $D_1$ is given by Eq.~(\ref{eq:D1}).
 
  There are several important differences with respect to the Gaussian RP model~\cite{facoetti,bogomolny} due to the fact that the self-energy $S_i$ is broadly distributed~\cite{}: 
In the Gaussian RP ensemble the width of the mini-bands 
is simply given by
the average effective spectral bandwidth $\Gamma_{\rm av} = \langle S_i(z) \rangle$ that a particle created in $i$ can reach, Eq.~(\ref{eq:mb}): 
$E_{\rm Th} \sim 
N \langle {\cal H}_{ij}^2 \rangle \sim N^{1-\gamma}$. As discussed above and illustrated in Fig.~\ref{fig:thouless}, for its L-RP counterpart 
one can in principle define a typical and an average bandwidth which exhibit a different scaling with $N$ for $\mu<2$:
\[
\begin{aligned}
\Gamma_{\rm typ} &\approx 2 \pi \rho N  [{\cal H}_{ij}^2]_{\rm typ} \sim N^{1 - 2 \gamma / \mu} \, , \\
\Gamma_{\rm av} & \approx 2 \pi \rho N  \langle {\cal H}_{ij}^2 \rangle_W \sim N^{1 - \gamma} \, .
\end{aligned}
\]
These energy scales are both different from the Thouless energy, $\Gamma_{\rm typ}  \ll E_{\rm Th} = N^{(1 - \gamma )/ (\mu - 1)} \ll \Gamma_{\rm av}$, which is instead determined self-consistently as the typical value of the self energy which sets the scale at which the spectral statistics exhibits a crossover. 
Note that the scaling of $\Gamma_{\rm typ}$, $\Gamma_{\rm av}$, and $E_{\rm Th}$ all coincide and become equal to the ones of the Gaussian RP ensemble $N^{1 - \gamma}$ for $\mu \to 2$. Therefore for $\mu \ge 2$ the mini-bands in the local spectrum become {\it fractal} (and not multifractal) and the anomalous dimensions become degenerate, $D_q = D$ for $q > 1/2$~\cite{kravtsov}.

\section{Numerical study of the Fractal dimensions} \label{sec:D}
 
 	\begin{figure*}
 		%
 		%
 		\hspace{-0.5cm} \includegraphics[width=0.99\textwidth]{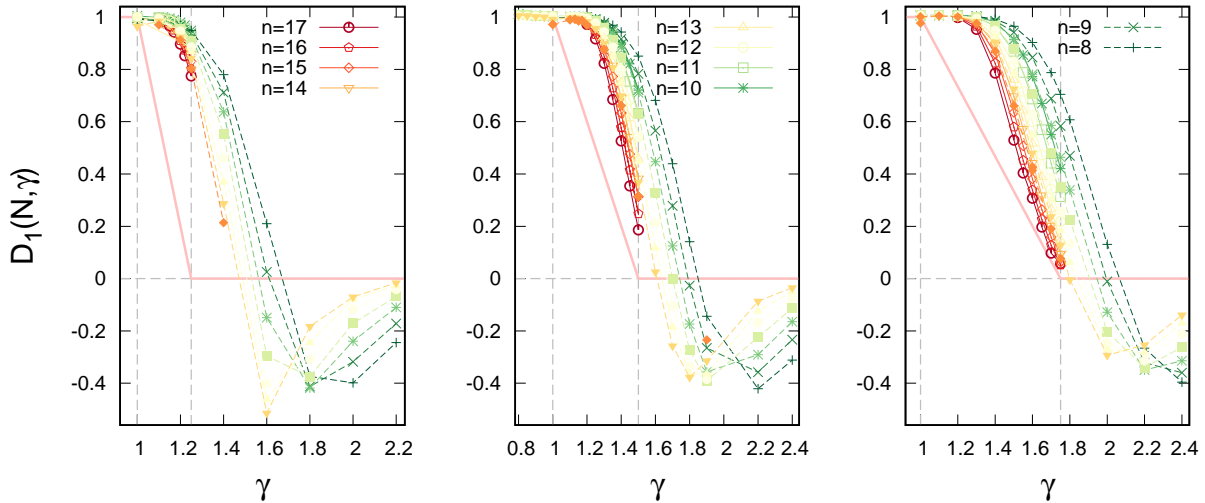}
 		\vspace{-6cm}
 		\caption{\label{fig:Df} Flowing $N$-dependent fractal dimensions $D_1 (N,\gamma)$ 
 			for $\mu=1.25$ (left),  $\mu=1.5$ (middle),  $\mu=1.75$ (right),  and $N=2^n$ with $n=8,\ldots,17$.  The pink lines corresponds to the analytic prediction~(\ref{eq:D1}). $D_1(N,\gamma)$ is computed either from the numerical solution of the cavity equations~(\ref{eq:cavity}) and using Eq.~(\ref{eq:D1cavity}) (continuous curves, for $10\le n \le 17$), or via Eq.~(\ref{eq:D1ed}) from the scaling  of the first moment of the wave-functions' amplitudes measured from exact diagonalizations (dashed curves, for $8\le n \le 15$). The results obtained using these two procedures are essentially indistinguishable within the numerical accuracy for all values of $N$, $\mu$, and $\gamma$ considered. }
 	\end{figure*}
 
 The analytical predictions~(\ref{eq:D1}) and~(\ref{eq:Dinf})
and can be directly checked by the analysis of the finite size scaling behavior of 
the flowing fractal dimension $D_1(N,\gamma)$ and $D_\infty(N,\gamma)$, which can be  measured either from the full numerical solution of the cavity equations~(\ref{eq:cavity}) or from exact diagonalizations. 
Solving Eqs.~(\ref{eq:cavity}) for  $N \times N$ matrices of the L-RP ensemble, and averaging the results over several independent realizations of the disorder, one can estimate $D_1(N,\gamma)$ from the  derivative of  the logarithm of $\rho_{\rm typ} (N,\gamma)$ with respect to $\log N$:
\begin{equation} \label{eq:D1cavity}
D_1 (N,\gamma) = 1 + \frac{\partial \log \rho_{\rm typ} (N,\gamma)} {\partial \log N} \, .
\end{equation}
(Hereafter the logarithmic derivatives 
are 
computed as discrete derivatives involving the three available values of the system size closest to $N$.)
Similarly, $D_1(N,\gamma)$ can be also computed via exact diagonalizations from the scaling behavior of the first moment of the wave-functions amplitudes with the system size:
\begin{equation} \label{eq:D1ed}
\begin{aligned} 
&\Upsilon_1 (n) 
= - 
\sum_{i=1}^N | \psi_n (i) |^2 \log \left ( | \psi_n (i) |^2 \right) 
\, , \\
&D_1 (N,\gamma)  =  \frac{\partial \log \langle  \Upsilon_1  (N,\gamma) \rangle} {\partial \log N} \, .  
\end{aligned}
\end{equation}
(The averages $\langle  \Upsilon_1  (N,\gamma) \rangle$ are computed over different realizations of the disorder and over eigenvectors within a given energy band around the middle of the spectrum, e.g., $E_n \in [-W/4,W/4]$.)
The numerical results obtained using these two procedures are shown in 
Fig.~\ref{fig:Df} as a function of $\gamma$ for three values of $\mu$ and for several values of the system size. The figure illustrates  that the estimations for  $D_1(N,\gamma)$ obtained from the cavity approach and from exact diagonalizations are essentially indistinguishable within the numerical incertitudes (see also Fig.~\ref{fig:stab}), which is not surprising since the cavity equations are asymptotically exact for the L-RP ensemble at large $N$.
The quasi-plateau of $D_1 (N,\gamma)$ observed close to the ergodic transition, $\gamma \gtrsim 1$, is a manifestation of the fact that the line of critical points separating the delocalized regime from the NEE one is in the ergodic phase.

\begin{figure}
	 \includegraphics[width=0.49\textwidth]{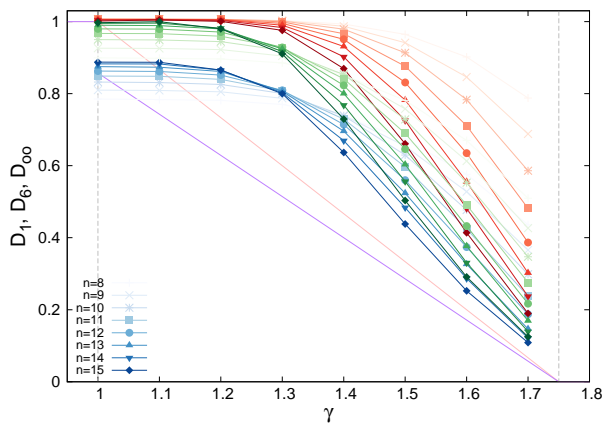}
	 	\vspace{-1cm}
	\caption{\label{fig:DfinfD6} Flowing $N$-dependent fractal dimensions $D_1 (N,\gamma)$ (red),  $D_6 (N,\gamma)$ (green), and $D_\infty  (N,\gamma)$ (blue) for $\mu=1.75$ in the NEE phase ($1 < \gamma < \mu$) for
		  $N=2^n$ with $n=8,\ldots,15$.  The fractal dimensions are estimated via Eqs.~(\ref{eq:D1ed}) and~(\ref{eq:D2ed}).  The pink line corresponds to the analytic prediction for $D_1$, Eq.~(\ref{eq:D1}), while the violet line shows the analytic prediction for $D_{\infty}$, Eq.~(\ref{eq:Dinf}) .}
\end{figure}

From exact diagonalizations one can also measure higher moments of the wave-functions amplitudes: 
\begin{equation} \label{eq:D2ed}
\begin{aligned} 
&\Upsilon_q (n) 
= 
\log \left (\sum_{i=1}^N  | \psi_n (i) |^{2q} \right) 
\, ,  \\
&(q-1) D_q (N,\gamma) =  - \frac{\partial \log \langle  \Upsilon_q  (N,\gamma) \rangle} {\partial \log N} \, .  
\end{aligned}
\end{equation}
The flowing fractal exponent $D_2 (N, \gamma)$, associated with the scaling with $N$ of the inverse participation ratio (IPR) is  plotted in Fig.~\ref{fig:D2} of App.~\ref{app:D}, showing that $D_2 (N, \gamma)$ has the same qualitative behavior of (and  is slightly smaller than) $D_1 (N, \gamma)$. 
In Fig.~\ref{fig:DfinfD6} we plot $D_q (N,\gamma)$ for $q=1$ (red), $q=6$ (green), and $q \to \infty$ (blue),  for $\mu=1.75$ within the NEE phase, showing that  $D_\infty < D_6 < D_1$.

\begin{figure}
	\includegraphics[width=0.49\textwidth]{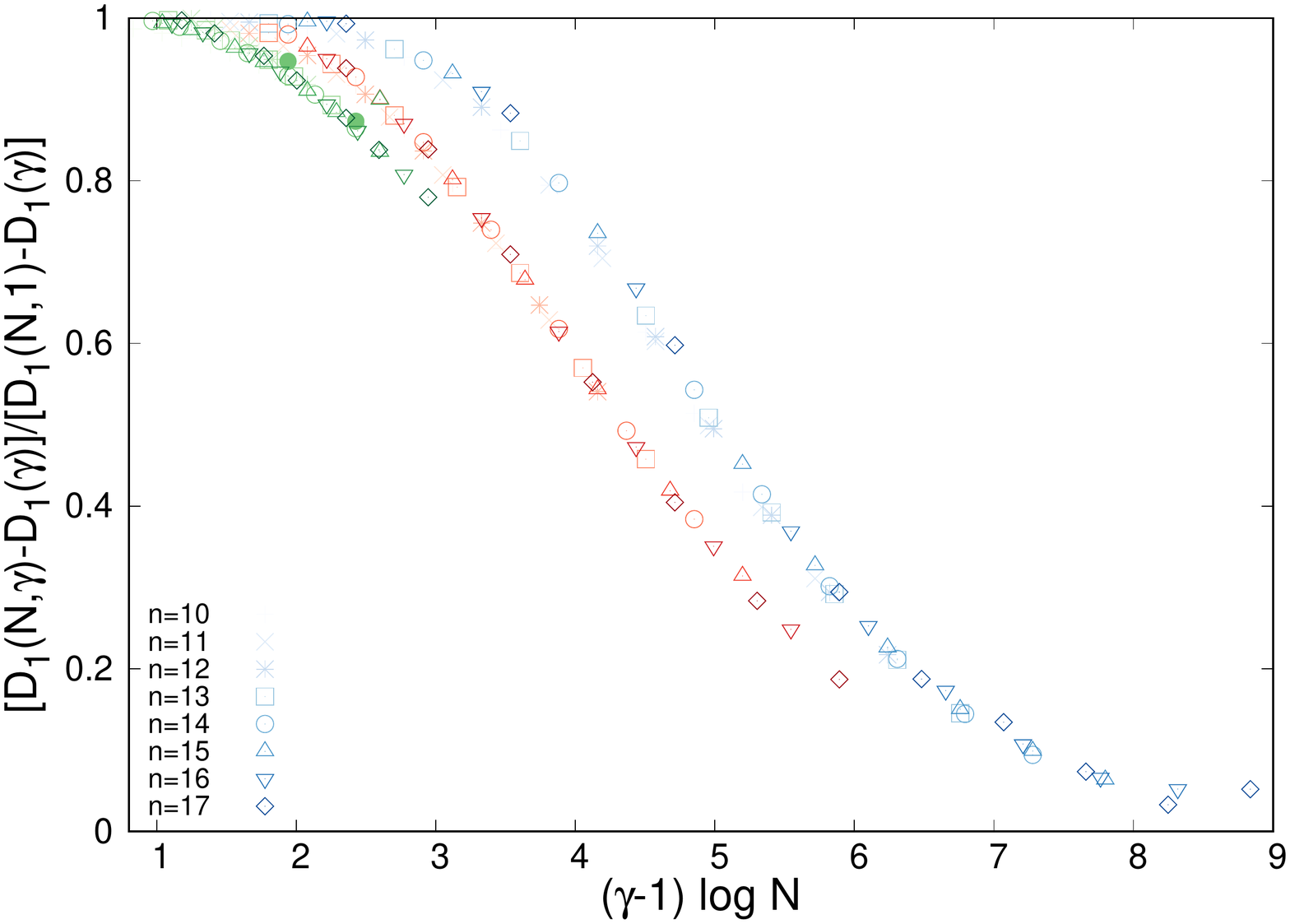}
	
	\vspace{-1cm}
	\includegraphics[width=0.49\textwidth]{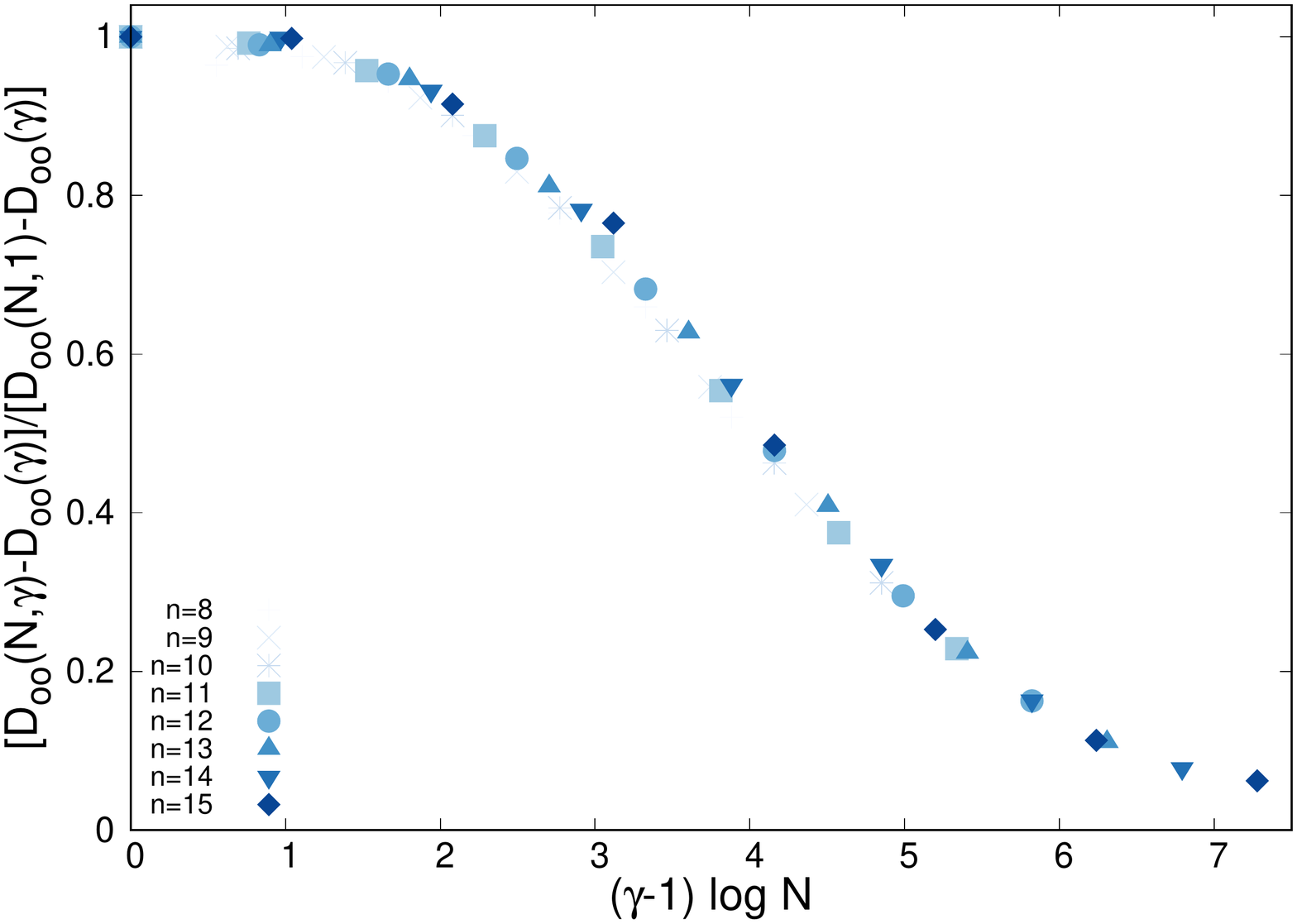}
		\vspace{-1cm}
	\caption{\label{fig:Dall} Finite-size scaling analysis of the distance of the flowing $N$-dependent fractal dimensions $D_1 (N,\gamma)$ (top) and $D_\infty (N,\gamma)$ (bottom)  from their theoretical asymptotic values, Eqs.~(\ref{eq:D1}) and~(\ref{eq:Dinf}), divided by the same quantities at small $N$, for $\mu=1.25$ (green),  $\mu=1.5$ (red),  $\mu=1.75$ (blue). 
		A very good data collapse is obtained for all values of $\mu$ when the ratios  $[D_1(N,\gamma) - D_1(\gamma)]/[D_1(N,1) - D_1(\gamma)]$ and $[D_\infty(N,\gamma) - D_\infty(\gamma)]/[D_\infty(N,1) - D_\infty(\gamma)]$ 
		are plotted as a function of the scaling variable $(\gamma - 1) \log N$.}
\end{figure}

In order to check that the $D_1(N,\gamma)$ and $D_\infty(N,\gamma)$  asymptotically approach the theoretical predictions in the large $N$ limit, 
we have performed a finite size scaling analysis of the distance between the flowing fractal exponents 
from their theoretical asymptotic value, Eqs.~(\ref{eq:D1}) and~(\ref{eq:Dinf}). In order to have that the data at different values of $N$ and $\gamma$ vary on the same scale (i.e., between $0$ and $1$), we have considered the ratio of $D_1(N,\gamma) - D_1(\gamma)$ [resp, $D_\infty(N,\gamma) - D_\infty(\gamma)$] divided by the amplitude of the same quantity at small $N$, $D_1(N,\gamma=1) - D_1(\gamma)$ [resp.  $D_\infty(N,\gamma=1) - D_\infty(\gamma)$]~\cite{gabriel,mace}. Figs.~\ref{fig:Dall} clearly show that a very good collapse is obtained for all values of $\mu$ when the data for $q=1$ and $q=\infty$ are plotted in terms of the scaling variable $(\gamma - \gamma_{\rm ergo}) (\log N)^{1/\nu_{\rm ergo}}$, with $\gamma_{\rm ergo} = 1$. The best collapse is found for $\nu_{\rm ergo}=1$ independently of $\mu$  (see Fig.~\ref{fig:nu}), as for the RP model~\cite{pino}.
This finite-size scaling analysis confirms that the fractal dimensions are not degenerate for $\mu<2$, as anticipated above as a direct consequence of the multifractality of the mini-bands.

As shown in Figs.~\ref{fig:D} and~\ref{fig:D126inf} of App.~\ref{app:D}, an independent estimation of 
$\nu_{\rm ergo}$ 
can be also obtained by performing a  finite size scaling analysis of the  moments of the wave-functions amplitudes with the system size similar to the one proposed in Refs.~\cite{mace} (and inspired by the analysis of Ref.~\cite{gabriel}) 
on the insulating side of  the MBL transition. This analysis confirms that $\nu_{\rm ergo}=1$ at the ergodic transition independently of $\mu$.

\section{Level statistics} \label{sec:level}

In this section we show  finite-size scaling analysis of the level statistics obtained from exact diagonalizations of L-RP random matrices of size $N = 2^n$ with $n$ ranging from $8$ to $15$. 
Averages are performed over different realizations of the disorder and over eigenstates within an energy window around the middle of the spectrum, $E_n \in [-W/4,W/4]$. 

\begin{figure}
	\includegraphics[width=0.49\textwidth]{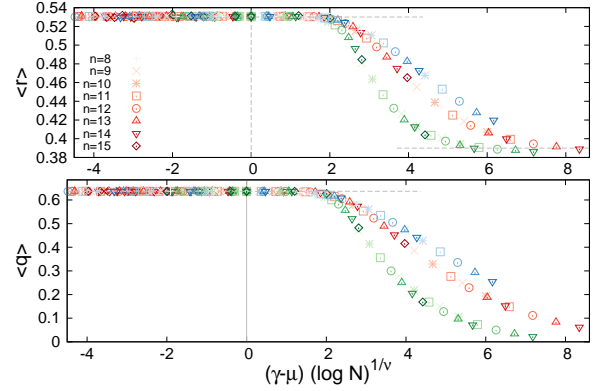}
		\vspace{-1cm}
	\caption{\label{fig:R} $\langle r \rangle$ (top) and $\langle q \rangle$ (bottom) as a function of $\gamma$ for $\mu = 1.25$ (green), $\mu = 1.5$ (red), and $\mu=1.75$ (blue) for several system sizes $N = 2^n$ with $n$ from $8$ to $15$ (different values of $n$ correspond to different symbols as indicated in the legend). A very good data collapse is obtained in terms of the scaling variable $(\gamma - \gamma_{\rm AL}) (\log N)^{1/\nu_{\rm AL}}$, with $\gamma_{\rm AL} = \mu$ and $\nu_{\rm AL}^{-1} \approx 0.99$ for $\mu=1.75$, $\nu_{\rm AL}^{-1} \approx 0.98$ for $\mu=1.5$ and $\nu_{\rm AL}^{-1} \approx 0.9$ for $\mu = 1.25$.}
\end{figure}

 We start by focusing on the level statistics of neighboring eigenvalues and measure the ratio of adjacent gaps:
\[
r_n = {\rm min} \left \{ \frac{E_{n+2} - E_{n+1}}{ E_{n+1} - E_n} , \frac{ E_{n+1} - E_{n}}{E_{n+2} - E_{n+1}}  \right \}  \, ,
\]
whose probability distribution displays a universal form depending on the level statistics, with $\langle r \rangle$ equal to $0.53$ in the GOE ensemble and to $0.39$ for Poisson statistics~\cite{huse}.

The transition from GOE to Poisson statistics can also be captured by correlations between adjacent eigenstates such as the mutual overlap between two subsequent eigenvectors, defined as
\[
q_n = \sum_{i=1}^N \left \vert \psi_n(i) \right \vert \left \vert \psi_{n+1}(i) \right \vert \, ,
\]
In the GOE phase $\langle q \rangle$ converges to $2/\pi$ (as expected for random vector on a $N$-dimensional sphere), while in the localized phase two successive eigenvector are typically peaked around  different sites and do not overlap and $\langle q \rangle \to 0$. 

\begin{figure}
	\includegraphics[width=0.49\textwidth]{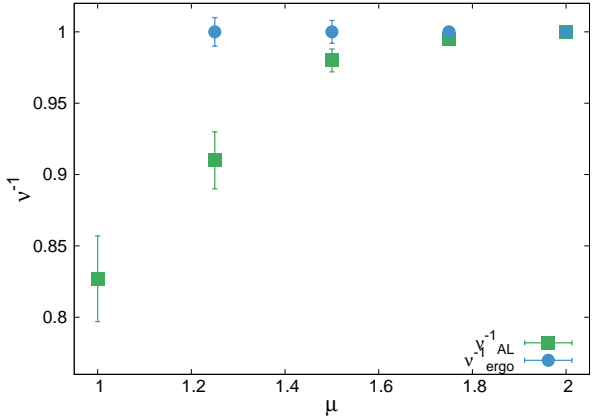}
		\vspace{-1cm}
	\caption{\label{fig:nu} Critical exponents $\nu_{\rm AL}$ and $\nu_{\rm ergo}$ yielding the best data collapse for the Anderson localization transition and the ergodicity breaking transition respectively.}
\end{figure}

In Fig.~\ref{fig:R} we plot $\langle r \rangle$ (top) and $\langle q \rangle$ (bottom) as a function of $\gamma$ for three values of $\mu \in (1,2)$, 
showing that a very good collapse is obtained for both observables when the data are plotted in terms of the scaling variable $(\gamma - \gamma_{\rm AL}) (\log N)^{1/\nu_{\rm AL}}$ (with $\gamma_{\rm AL} = \mu$), 
confirming that the level statistics of neighboring gaps exhibit a transition from GOE to Poisson at the AL transition (note that the critical point is in the GOE phase for all values of $\mu$).
The exponent $\nu_{\rm AL}$ that produces the best collapse is found to decrease continuously as $\mu$ is increased and approaches the RP value $\nu_{\rm AL}^{-1} = 1$ for $\mu=2$~\cite{pino} (see Fig.~\ref{fig:nu}).

\begin{figure}
	\includegraphics[width=0.49\textwidth]{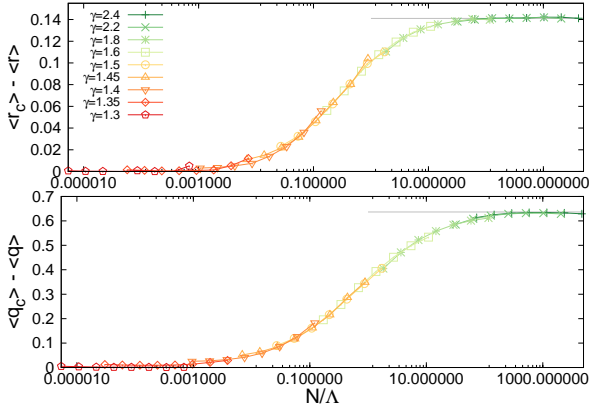}
		\vspace{-1cm}
	\caption{\label{fig:Rscaling1} $\langle r\rangle (1,N) - \langle r \rangle (\gamma,N)$ (top) and $\langle q \rangle (1,N) - \langle q \rangle (\gamma,N)$ (bottom) as a function of the scaling variable $N/\Lambda(\gamma)$. A good data collapse for both observables is obtained for $\Lambda(\gamma) \sim \exp [A (\gamma-1)^{- \nu_{\rm AL}}]$, with $\nu_{\rm AL}\approx 1.2$. The gray horizontal lines represent the difference between the GOE and Poisson asymptotic values.}
\end{figure}

For $\mu=1$ a reasonably good data collapse  of the observables $\langle r \rangle$ and $\langle q \rangle$ related to the statistics of neighboring gaps cannot be achieved by using $(\gamma - 1) (\log N)^{1/\nu_{\rm AL}}$ as a scaling variable for any value of $\nu_{\rm AL}$, and we have therefore attempted a different finite size scaling analysis, as illustrated in Fig.~\ref{fig:Rscaling1}. More specifically, we plot the distance of $\langle r \rangle (\gamma,N)$ and $\langle q \rangle (\gamma,N)$ from their values at the critical point ($\langle r \rangle_c \approx 0.53$ and $\langle q_c \rangle = 2/\pi$ for $\gamma=1$) as a function of the scaling variable $N/\Lambda (\gamma)$, where $\Lambda (\gamma)$ is a  correlation volume that depends on the distance from the critical point. 
A very good data collapse of both $\langle r \rangle$ and $\langle q \rangle$ is obtained for $\Lambda \sim \exp [A (\gamma-1)^{-\nu_{\rm AL}}]$ with $\nu_{\rm AL} \approx 1.2$. Such volumic scaling is similar to the critical scaling observed on the delocalized side of the Anderson model on the Bethe lattice~\cite{gabriel} and reflects the fact that at the tricritical point ($\mu=1$, $\gamma=1$) the fractal dimension exhibit a discontinuous jump from $D _1= 1$ for $\gamma = 1$ to $D_1 \to 0$ for $\gamma \to 1^+$. However in the present case the situation is somehow reversed compared to the Anderson model on the Bethe lattice, in the sense that here the critical point is in the delocalized phase (i.e., $D_1 = 1$ for $\mu = 1$ and the level statistics is GOE) and the scaling in terms of an exponentially large correlation volume is found on the {\it localized} side of the transition, while for the Anderson model on the Bethe lattice the critical point is in the localized phase (i.e., $D = 0$ at $W_c$ and the statistics is Poisson) and the volumic scaling is found on the {\it delocalized} side of the transition~\cite{gabriel,susy,tikhonov}.

\section{Overlap correlation function} \label{sec:K2}

It is also worthwhile to study the behavior of  
the spectral correlation function $K_2 (\omega)$ between eigenstates at different energy, which provides
 a very useful probe of the level statistics and on the statistics of wave-functions' amplitudes, and allows one to distinguish between ergodic, localized, and multifractal states~\cite{altshulerK2,mirlin,chalker,kravK2}:
\begin{equation} \label{eq:K2}
	\begin{aligned}
K_2 (\omega) & = \left \langle \sum_i \vert  \psi_n (i) \psi_m (i) \vert^2 \delta \left( E_n - E_m - \omega \right) \right \rangle \\
& \simeq \lim_{\eta \to 0^+} \left \langle \frac{N \sum_i{\rm Im} {\cal G}_{ii} (\omega/2) \, {\rm Im} {\cal G}_{ii} (-\omega/2)}{\sum_i{\rm Im} {\cal G}_{ii} (\omega/2) \sum_i {\rm Im} {\cal G}_{ii} (-\omega/2)} \right \rangle \, .
\end{aligned}
\end{equation}
Furthermore, $K_2 (\omega)$ is the Fourier transform of the return probability and can be thought as proxy for the correlation function of local operators, e.g. the spin-spin correlation function,  in the problem of many-body localization~\cite{serbyn,polk,dinamica,tikhK2}.

For GOE matrices $K_2(\omega) = 1$ identically, independently on $\omega$ on the entire spectral bandwidth. In the standard (ergodic) metallic phase (i.e., the ``weakly'' ergodic phase using the terminology of Ref.~\cite{khay}) $K_2(\omega)$ has a plateau at small energies, for $\omega < E_{\rm Th}$, followed by a fast-decay which is described by a power-law, 
with a system-dependent exponent~\cite{chalker}. 
The height of the plateau is larger than one, which implies an enhancement of correlations compared to the case of independently fluctuating Gaussian wave-functions. The Thouless energy which separates the plateau from the power-law decay stays finite in the thermodynamic limit and extends to larger energies as one goes deeply into the metallic phase, and corresponds to the  energy band over 
which GOE-like correlations establish~\cite{altshulerK2}.

\begin{figure*}
	\includegraphics[width=0.49\textwidth]{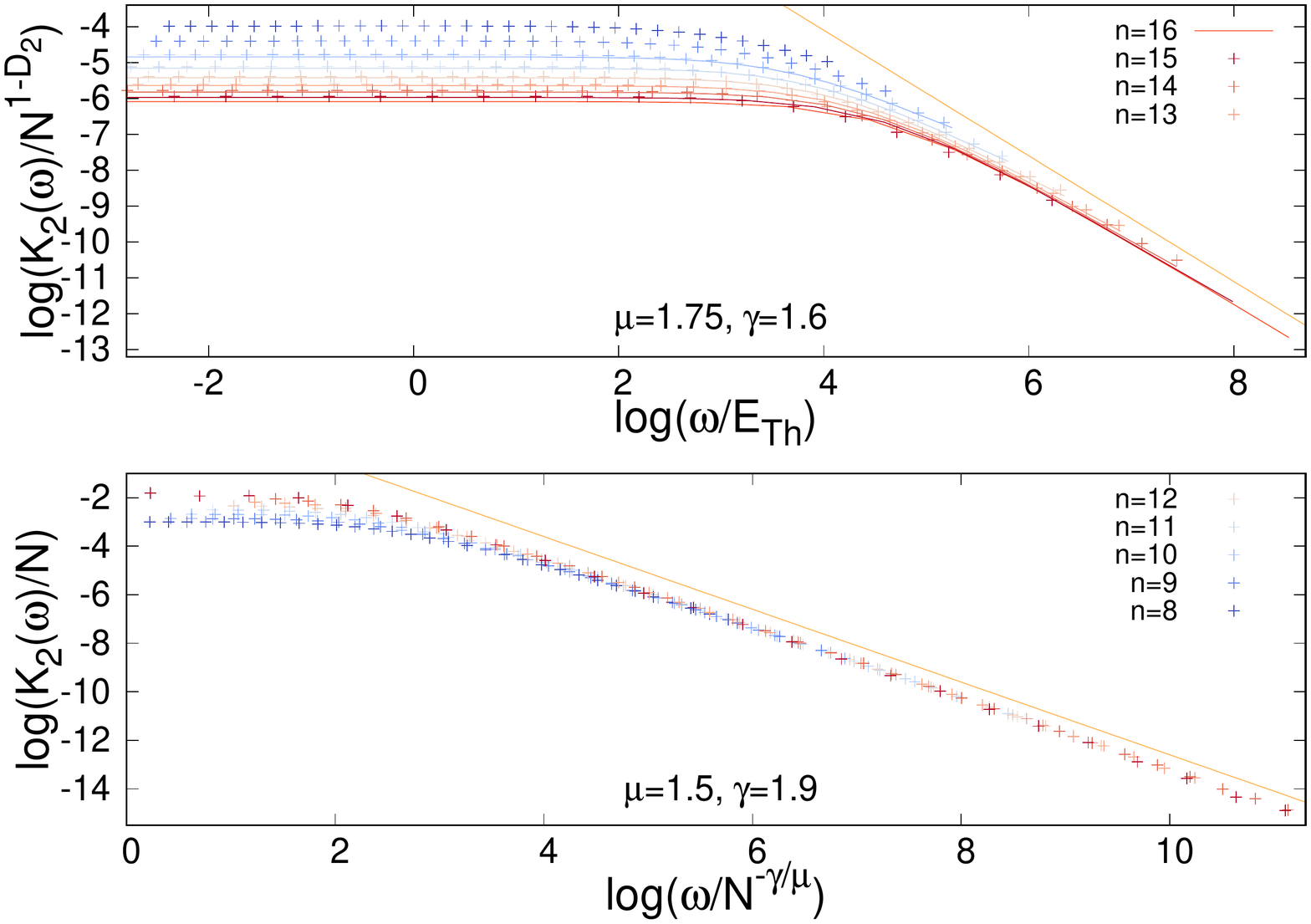}
	\includegraphics[width=0.49\textwidth]{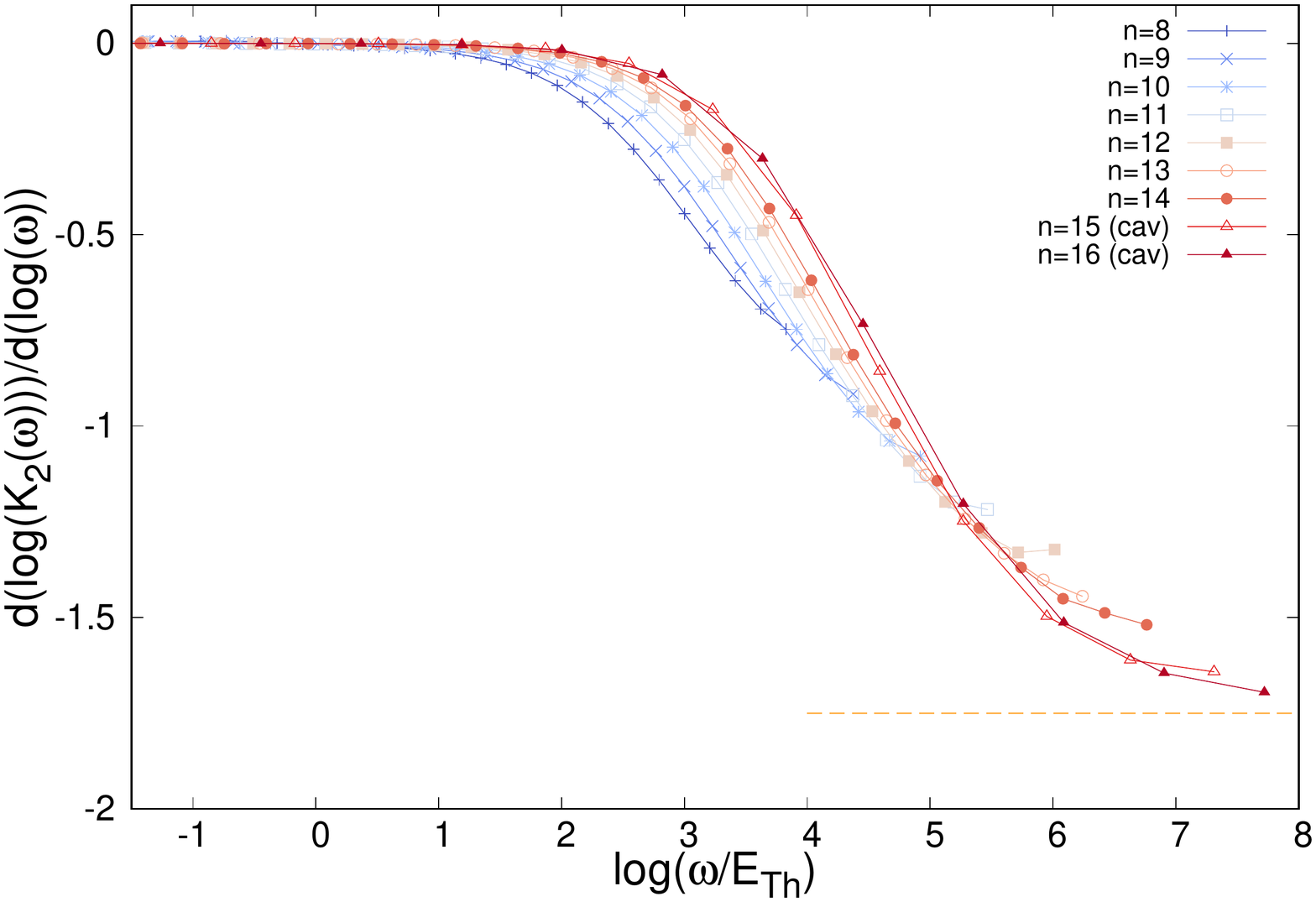}
		\vspace{-1cm}
	\caption{\label{fig:K} Top left: Logarithm of  the overlap correlation function versus $\log \omega$ in the NEE phase,  $\mu = 1.75$ and $\gamma=1.6$, for several system sizes $N=2^n$, with $n=8, \ldots, 16$ (different colors correspond to different value of $n$). The results obtained from exact diagonalizations (for $8 \le n \le 15$) are represented with points, while the results obtained using the cavity approach (for $10 \le n \le 16$) are represented with full  lines. 
		The energy axis is rescaled by the Thouless energy $E_{\rm Th} = N^{(1-\gamma)/(\mu-1)}$, while the vertical axis is rescaled by $N^{1-D_2}$ (with $D_2 \approx 0.19 \lesssim D_1 = 0.2$), i.e., the value of $K_2 (\omega)$ for $\omega \to 0$. Bottom left:  Logarithm of  the overlap correlation function as a function of $\log \omega$ in the AL phase,  $\mu = 1.5$ and $\gamma=1.9$, for several system sizes. The energy axis is rescaled by $N^{-\gamma/\mu}$, while the vertical axis is rescaled by $N$. The plateau at small energy is followed in both phases by a power-law decay $K_2(\omega) \propto 1/\omega^\mu$  (orange lines). Right:  Derivative
			$\partial \log K_2 (\omega) / \partial \log \omega$ as a function of $\log (\omega / E_{\rm Th})$ for $\mu=1.75$, $\gamma=1.6$, and for several system sizes $N=2^n$, which gives a running with $\omega$ and $N$ exponent $\theta$ of a local power-law describing the decay of $K_2 (\omega)$. $\theta=0$ for $\omega < E_{\rm Th}$ (as in the Gaussian RP and in LN-RP models), while for $\omega \gg E_{\rm Th}$ the exponent $\theta$ tends to $\mu$ at large $N$, corresponding to a non-trivial fractal structure of the set of mini-bands. Data for $8 \le n \le 14$ have been obtained from exact diagonalizations while data for $n=15,16$ have been obtained using the cavity approach.}
\end{figure*}

The behavior of the overlap correlation function for
multifractal wave-functions is instead drastically different: In the NEE phase of the Gaussian RP ensemble, for instance, the plateau is present only in a
narrow energy interval, as $E_{\rm Th}$ shrinks
to zero in the thermodynamic limit (still staying much larger than the mean level spacing), while its height grows as $N^{1-D_2}$.
Beyond $E_{\rm Th}$  eigenfunctions poorly overlap with each other and the statistics is no longer GOE and $K_2(\omega)$ decay to zero as a power-law, $K_2 (\omega) \sim (\omega/E_{\rm Th})^{-2}$~\cite{kravtsov}.

Our numerical results are presented in Fig.~\ref{fig:K}. 
The overlap correlation function is computed using both exact diagonalizations and its spectral representation in terms of the Green's functions obtained via the cavity method~\cite{footnote2}, finding a very good agreement between the two approaches.   In the top panel we plot $K_2(\omega)$ for several system sizes in the NEE phase, $\mu = 1.75$ and $\gamma=1.6$, and we show that in the large $N$ limit the data corresponding to different sizes approach a limiting curve when the energy is rescaled by the Thouless energy $E_{\rm Th} = N^{(1-\gamma)/(\mu-1)}$ and the vertical axis is rescaled by $N^{1-D_2}$, where $D_2 \lesssim D_1$. (A similar behavior is observed for other values of $\mu$ and $\gamma$ within the multifractal phase.) 
The fact that $K_2 (\omega)$ is constant for $N^{-1} < \omega < E_{\rm Th}$ reflects the fact that the mini-bands are locally compact, as in the Gaussian RP model (i.e., the fractal dimension  of the local spectrum {\it inside} a mini-band is equal to $1$). 
	At larger energy separation, $\omega \gg E_{\rm Th}$,  the exponent $\theta = - \partial \log K_2 (\omega) / \partial \log \omega$ 
	reflects instead the fractal structure of the set of mini-bands~\cite{khay}. For the Gaussian RP $\theta = 2$~\cite{kravtsov,facoetti,bera}, while, as shown in the right panel of Fig.~\ref{fig:K}, $\theta$ tends to $\mu$ at large $N$ in the NEE regime, $\mu \in (1,2)$ and $\gamma \in (1,\mu)$, of the L-RP ensemble,  irrespectively of the value of $\gamma$.
	
In the bottom panel of Fig.~\ref{fig:K} we show the results in the AL phase, $\mu = 1.5$ and $\gamma=1.9$. In this case $K_2(\omega)$ displays the usual localized behavior, despite the fact that for $\gamma<2$ the average effective bandwidth $\Gamma_{\rm av} \sim N \langle {\cal H}_{ij}^2 \rangle_W \sim N^{1-\gamma}$ is still much larger than the mean level spacing $N^{-1}$. A good collapse of the data at different $N$ is obtained when the vertical axis is rescaled by $N$  ($D_2=0$) and the energy axis is rescaled by $N^{-\gamma/\mu}$. The  plateau that extends up to an energy scale $N^{-\gamma/\mu} < N^{-1}$  corresponds to rare resonances when $\omega < | {\cal H}_{ij}|_{\rm typ}$.
Also in the AL regime,  the plateau at small energy is followed by a fast decrease $K_2(\omega) \propto 1/\omega^{\mu}$ (orange line).

\section{Conclusions} \label{sec:conclusions}

In this paper we have studied a generalization of the RP ensemble when the off-diagonal perturbation belongs to the L\'evy universality class~\cite{monthus-LRP}, with i.i.d. matrix elements with power law tails of exponent $1 + \mu$ and typical value scaling as $N^{-\gamma / \mu}$. 
We believe that the L-RP ensemble provides a more realistic benchmark to develop an effective description of delocalization of the wave-functions in interacting many-body disordered systems, in which  
the effective transition rates between distant states in the Hilbert space  correspond to a long series of quantum transitions and are in general  broadly distributed~\cite{tarzia,qrem3,war,roy}.

The most important feature of  the model is that, 
due to the fat tails of  the off-diagonal matrix elements, sites at energy separation much larger than the typical bandwidth $N [{\cal H}_{ij}^2]_{\rm typ}$ can be hybridized by anomalously large rare matrix elements, producing a NEE phase with multifractal mini-bands.
In this sense the L-RP ensemble is much richer than its Gaussian RP counterpart, since the mini-bands in the local spectrum are  multifractal and the spectrum of fractal dimension is not degenerate.

One of the most important outcome of our analysis is the formulation of a new, simple, intuitive, and physically transparent argument that allows one to characterize the multifractal structure of the mini-bands and determine the fractal dimensions of the eigenstates in the NEE phase, as well as the phase diagram of the system.
The basic idea is that the Thouless energy can be determined self-consistently by imposing that hybridization occurs provided that the largest matrix elements between a site $i$ and the other $N^{D_1}$ sites $j$ within a given mini-band are of the same order of the energy spreading of the mini-bands itself. This argument is very general and can be in principle extended and adapted to analyze the multifractal states also in more complex situations in which, for instance, the effective transition rates are correlated~\cite{nosov} and/or depend on the positions $i$ and $j$ in the reference space and on the energy separation $|\epsilon_i - \epsilon_j|$, as in many-body problems~\cite{qrem3,tarzia,war,roy}. Extending our analysis to these situations is certainly a promising direction for future investigations.

The predictions of such simple arguments are fully confirmed both analytically, by a thorough analysis of the self-consistent equations for the diagonal elements of the resolvent matrix obtained using the cavity approach, 
and numerically, by means of extensive exact diagonalizations, and are also in full agreement with the ``rules of thumb'' criteria for localization and ergodicity recently put forward in Refs.~\cite{kravtsov1,nosov,bogomolny,khay}.

Another interesting feature of the model, is the existence of a tricritical point~\cite{kravtsov1,khay} for $\mu = 1$ (i.e., Cauchy distributed off-diagonal elements~\cite{cauchy}) and $\gamma=1$,
 where the fractal dimensions exhibit a discontinuous jump from $D _1= 1$ for $\gamma = 1$ to $D_1 \to 0$ for $\gamma \to 1^+$. This is somehow a specular behavior compared to the  Anderson model on the Bethe lattice: here the tricritical point is in the delocalized phase (i.e., $D_1 = 1$ for $\mu = 1$ and the level statistics is GOE) and the scaling in terms of an exponentially large correlation volume is found on the {\it localized} side of the transition, while for the Anderson model on the Bethe lattice the critical point is in the localized phase (i.e., $D = 0$ at $W_c$ and the statistics is Poisson) and the volume scaling is found on the {\it delocalized} side of the transition~\cite{gabriel,susy,tikhonov}.

\begin{acknowledgments}
	We would like to thank  I. M. Khaymovich and V. E. Kravtsov for many enlightening and helpful discussions. 
	This work was partially supported by the grant from the Simons Foundation (\#454935 Giulio Biroli). 
\end{acknowledgments}

\appendix

\section{$\mu \in (0,1)$} \label{sec:mu1}

As discussed in Sec.~\ref{sec:PD}, for $\mu<1$ the L\'evy-RP ensemble exhibits a single discontinuous transition at $\gamma=1$ between a 
phase for $\gamma>1$ in which the off-diagonal L\'evy matrix elements are a small  regular perturbation and ${\cal H}$ is close to ${\cal A}$ (and eigenvectors are fully localized) to a phase for $\gamma<1$ in which the off-diagonal matrix elements dominate and ${\cal H}$ is close to ${\cal L}$. This is confirmed by the numerical results (not shown) that indeed clearly indicate that the level statistics tends to Poisson for $\gamma>1$. However from the analysis of  Ref.~\cite{noi} we know that L\'evy matrices have a mobility edge which separates an extended  phase at low energy from a  Anderson localized phase at high energy. For the natural scaling $\gamma=1$, when the typical value of ${\cal L}_{ij}$ is of order $N^{-1/\mu}$ (and the eigenvalues of  ${\cal L}$ are of order 1), the mobility edge is found at a finite energy $E_{\rm loc} (\mu)$ (which can be computed analytically~\cite{noi}).  $E_{\rm loc} (\mu)$ goes to $0$ for $\mu \to 0$ and to $+\infty$ for $\mu \to 1^-$. In other words, the fraction of extended and localized eigenstates of the spectrum are both extensive for $\mu \in (0,1)$; the fraction of extended states vanishes for $\mu \to 0$ while the fraction of localized states vanishes for $\mu \to 1^-$. When $\gamma<1$ the eigenvalues of ${\cal L}$ are all rescaled by $N^{(1-\gamma)/\mu}$ and the mobility edge is thus found at energy $E_{\rm loc} (\mu) N^{(1-\gamma)/\mu}$. We then expect that the phase transition taking place at $\gamma=1$ is in fact split in two: At low energy, $E<E_{\rm loc} (\mu) N^{(1-\gamma)/\mu}$, one has a discontinuous phase transition from the extended phase of L\'evy matrices for $\gamma<1$ to a AL phase dominated by the diagonal disorder for $\gamma>1$; At high energy,  $E>E_{\rm loc} (\mu) N^{(1-\gamma)/\mu}$, instead, one has a discontinuous transition from two different localized phases, namely a phase for $\gamma<1$ where eigenstates are close to the AL eigenstates of the off-diagonal L\'evy matrix, to a phase for $\gamma>1$ in which the eigenstates are localized due to the diagonal disorder.

\begin{figure}
	\includegraphics[width=0.49\textwidth]{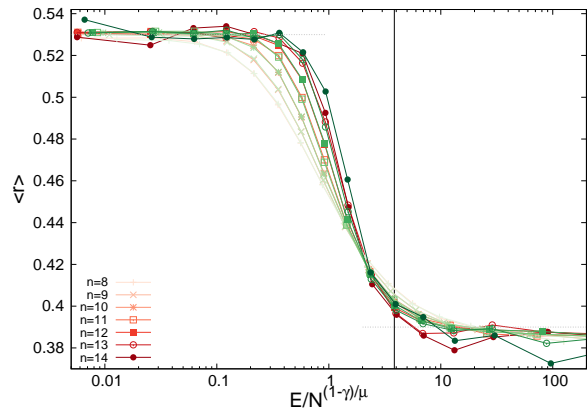}
	\caption{\label{fig:R05} $\langle r \rangle$ as a function of the energy (rescaled by $N^{(\gamma-1)/\mu}$ in order to have a $N$-independent spectrum in the thermodynamic limit)  for $\mu = 0.5$, for several system sizes $N = 2^n$ with $n$ from $8$ to $14$ (different values of $n$ correspond to different symbols as indicated in the legend) and for $\gamma=0$ (red) and $\gamma=0.8$ (green). The data indicated the presence of a transition from GOE statistics to Poisson statistics when the energy is increased above $E_{\rm loc}$, as already studied in Ref.~\cite{noi}. The curves at different values of $\gamma$ are essentially indistinguishable within numerical errors, implying the fact that the diagonal energies play no role. The vertical black line show the position of the mobility edge $E_{\rm loc} \approx 3.85$ computed analytically in~\cite{noi} for $W=0$.}
\end{figure}

This scenario  is fully confirmed by the numerical results of Fig.~\ref{fig:R05}, where we plot  $\langle r \rangle$ as a function of the energy (rescaled by $N^{(\gamma-1)/\mu}$ in order to have energies of order $1$)  for $\mu = 0.5$, for several system sizes, and for two values of $\gamma$. The curves indicate the presence of a transition from GOE statistics to Poisson statistics when the energy is increased above $E_{\rm loc}$, as already studied in Ref.~\cite{noi}. The transition point  $E_{\rm loc}$ in the rescaled variables does not depend on $\gamma$ in the whole interval $\gamma \in (0,1)$. 

\section{Stability of non-ergodic states against hybridization} \label{sec:stability}

In this section we discuss the stability criterion of non-ergodic  states against hybridization put forward in Ref.~\cite{khay} for the LN-RP ensemble.
Let us consider two states $\psi_n$ and $\psi_m$ on different fractal support sets. Let us assume that both states are multifractal and occupy $N^{D_1}$ sites of a support set where $| \psi (i) |^2 \sim N^{-D_1}$.

\begin{figure}
	\includegraphics[width=0.49\textwidth]{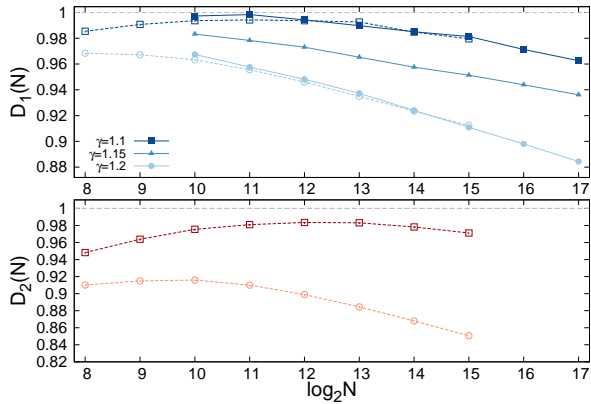}
	\vspace{-1cm}
	\caption{\label{fig:stab} Flowing $N$-dependent fractal dimensions $D_1 (N,\gamma)$ (top) and $D_2 (N,\gamma)$ (bottom) as a function of  $\log_2 N$ 
		computed either from the numerical solution of the cavity equations and using Eq.~(\ref{eq:D1cavity}) (continuous curves and filled symbols, for $10\le n \le 17$), or via Eqs.~(\ref{eq:D1ed}) and~(\ref{eq:D2ed}) from the scaling  of the first and second moments of the wave-functions amplitudes measured from exact diagonalizations (dashed curves and empty symbols, for $8\le n \le 15$).
		The value of $\mu=1.25$  is chosen in the middle of the region in which the bound~(\ref{eq:bounds}) with $\zeta=1/2$ would predict that  the NEE phase is unstable against hybridization, while the numerical data clearly shows that the anomalous dimensions are smaller than $1$ in the large $N$ limit.}
\end{figure}

We now  apply the usual Mott's argument for hybridization of states when the disorder realization changes from ${\cal L}_{ij}$ to ${\cal L}_{ij}^\prime$. 
The  new idea of Ref.~\cite{khay} is to compute the hopping matrix element ${\cal V}_{n m}$ between the states and not between the sites as is customary:
\[
{\cal V}_{n m} = \sum_{i,j} \delta {\cal L}_{ij} \psi_n(i) \psi_m(j) \, ,
\]
where $\psi_n(i)$   is the eigenfunction of the $n$-th state of ${\cal H}$. 
$\delta {\cal L}_{ij} = {\cal L}_{ij} - {\cal L}^\prime_{ij}$, where ${\cal  L}^\prime_{ij}$ is drawn from the same L\'evy distribution as   ${\cal L}_{ij}$, and are L\'evy distributed random variables with power-law tails of exponent $1+\mu$ and typical value $2^{1/\mu} N^{-\gamma/\mu}$. For $\mu<2$ we can thus use the generalized central limit theorem for the sum of heavy-tailed distributed random variables from which we get that ${\cal V}_{n m}$ are also L\'evy distributed  with power-law tails with exponent $1+\mu$ and typical value:
\[
[{\cal V}_{n m}]_{\rm typ} = \left [ \frac{2}{N^\gamma } \sum_{i,j} | \psi_n(i) |^\mu | \psi_m(j) |^\mu \right ]^{1/\mu} \, . 
\]
The moments of the wave-functions' amplitudes give by definition $N \langle | \psi_n(i) |^\mu \rangle \sim N^{-D_{\mu/2} (\mu/2 - 1)}$. Assuming that wave-functions belonging to different mini-bands are not correlated, we 
have that:
\[
[ {\cal V}_{n m}]_{\rm typ} \sim N^{-\frac{\gamma }{ \mu} + D_{\mu/2} \left(\frac{2}{\mu }- 1 \right)} \, .
\]
The condition of stability of the multifractal phase against
hybridization is derived similar to the Anderson criteria of
stability, Eq.~(\ref{eq:AL}), of the localized states. The difference is that now we
have to replace the matrix element between the resonant sites
${\cal L}_{ij}$ by the matrix element ${\cal V}_{n m}$ between the resonant non-ergodic
states and take into account that on each of  $N_S=N^{1-D_1}$
different support sets there are $N^{D_1}$ wave-functions which
belong to the same mini-band and thus are already in resonance
with each other. Therefore the total number of independent
states-candidates $N_H$ for hybridization with a given state  should
be smaller than the total number of states $N_S N^{D_1} = N$ and larger than the number of support sets $N_S=N^{1-D_1}$.
In full generality we posit below that $N_H \propto N^{1 - \zeta D_1}$, with $0 < \zeta < 1$.
In Ref.~\cite{khay} the authors chose 
to use the geometric mean $N_H \propto \sqrt{N N_S} = N^{1 - D_1/2}$, i.e., $\zeta = 1/2$.
For $1 < \mu < 2$ the Mott's criterion of stability of the multifractal phase reads in the limit $N \to \infty$ reads:
\[
N^{1 - \zeta D_1 }\int_0^W {\cal V} P({\cal V}) \, {\rm d} {\cal V} \sim N^{1 - \zeta D_1 -\frac{\gamma}{\mu} + D_{\mu/2} \left(\frac{2}{\mu }- 1 \right)} < \infty \, .
\]
For the RP model the fractal dimensions are degenerate for $q>1/2$~\cite{kravtsov}. 
This is  not the case for the L-RP ensemble, due to the fact that the mini-bands are multifractal as clearly illustrated by Fig.~\ref{fig:DfinfD6}. Yet, since $D_q$ is  a decreasing function of $q$, for $1<\mu<2$ one can assume that $D_{\mu/2}$ is well approximated by  $D_1$.
This results in an upper bound 
for the stability of the multifractal phase of the form: 
\begin{equation} \label{eq:bounds}
	D_1 \left ( \zeta + 1 - \frac{2}{\mu} \right) \ge 1 - \frac{\gamma}{\mu} \, .
\end{equation}
Of course, this condition cannot be fullfilled if $\zeta < 2/\mu - 1$ since the left hand side becomes negative. This implies that if one chooses $\zeta=1/2$ as in Ref.~\cite{khay} one would conclude that the NEE phase is unstable against hybridization  in the interval $\mu \in (1,4/3)$ at least. In fact plugging the expression~(\ref{eq:D1}) for $D_1$ into Eq.~(\ref{eq:bounds}) one finds that for $\zeta=1/2$ the stability criterion is never satisfied except at the RP limit $\mu = 2$. Yet this is in strong disagreement with the numerical results on the flowing fractal exponents discussed in the previous section (see Figs.~\ref{fig:Df},~\ref{fig:DfinfD6}, and~\ref{fig:Dall}), as  further illustrated in Fig.~\ref{fig:stab} for $\mu=1.25$, deep in the region where the instability should supposedly take place. This plot shows that $D_1(N,\gamma)$ and $D_2(N,\gamma)$ have a non-monotonic behavior as a function of $N$ on a characteristic  scale that increases as $\gamma$ is decreased,  indicating that in the $N\to \infty$ limit $D_1$ and $D_2$ approach a value strictly smaller than $1$, as expected for a genuine multifractal phase.

A possible way out from this issue is  obtained  by positing that $\zeta$  depends on $\mu$ in such a way that the instability is avoided.
In fact, assuming that $\zeta > 2/\mu - 1$ and using Eq.~(\ref{eq:D1}), one obtains that the stability criterion~(\ref{eq:bounds})  can be fullfilled in the whole NEE phase provided that $\zeta > 1/\mu$. Moreover, for $\zeta = 1/\mu$ the bound~(\ref{eq:bounds}) is saturated, as suggested in Ref.~\cite{khay}.

\section{Comparison with the LN-RP ensemble} \label{app:LNRP}

In this section we illustrate how the simple argument put forward in Sec.~\ref{sec:D} to determine the effective width of the mini-bands (i.e., the Thouless energy) and their multifractal structure allows one to obtain the phase diagram and the  fractal exponent $D_1$ of the LN-RP ensemble recently introduced and studied in Refs.~\cite{kravtsov1,khay}.

The LN-RP model is a modification of the RP random matrix ensemble~\cite{kravtsov} in which the i.i.d. off-diagonal elements are taken from a log-normal law:
\begin{equation} \label{eq:LN}
	P({\cal H}_{ij}) = \frac{\exp \Big[ - \frac{\log^2 \left( N^{\gamma/2} {\cal H}_{ij} \right) }{p \gamma \log N} \Big ]}{\sqrt{\pi p \gamma \log N} \, |{\cal H}_{ij} |} \, , 
\end{equation}
in such a way that the typical value of ${\cal H}_{ij}$ is $N^{-\gamma/2}$.
The standard RP ensemble is recovered for $p \to 0$.

We now apply the argument illustrated in Sec.~\ref{sec:D} to this model.
Let us assume that in the NEE phase the mini-bands extend over $N^{D_1}$ adjacent energy levels. A site $i$ within a given mini-band can hybridize with the other sites $j$ of the same mini-band provided that the maximum of the $N^{D_1}$ hybridization rates ${\cal H}_{ij}$ is of the order of the width of the mini-band itself, $N^{D_1 - 1}$.
We introduce the exponent $\alpha$ to parametrize the scaling of the 
maximum of $N^{D_1}$ i.i.d. elements extracted from the log-normal distribution~(\ref{eq:LN}):
\[
{\rm max}_{j=1,\ldots,N^{D_1}}  \{ {\cal H}_{ij} \} \sim N^{-\alpha} 
\]
(here we only consider the leading term and neglect corrections of order $\log  N$).
A simple extreme value statistics calculation yields
\[
\frac{(\gamma/2 - \alpha)^2}{p \gamma} = D_1 \, , \qquad 0 \le \alpha \le \gamma/2 \, .
\]
In fact this expression is correct only if the maximum is larger than the typical value of the matrix element, $\alpha < \gamma/2$. 
Moreover, to be in the NEE phase, in which the effective total bandwidth is dominated by the diagonal disorder and is of order $W$, we need to require that $\alpha > 0$.

Imposing the self-consistent condition 
$N^{-\alpha} \sim N^{D_1 - 1}$ yields a self-consistent equation for $\alpha$ whose solution is:
\begin{equation} \label{eq:alpha}
	\alpha (\gamma,p) = \frac{(1 - p) \gamma - \sqrt{(1 - p)^2 \gamma^2 - \gamma^2 + 4 p \gamma}}{2} \, .
\end{equation}
(The relevant solution  is the one with the minus sign since, as explained above, one has to require that  $\alpha < \gamma/2$).
The transition to the ergodic phase corresponds to the points where $D_1=1$ (i.e., $\alpha=0$), $\gamma_{\rm ergo} = 4 p$, while the AL transition occurs when the solution of Eq.~(\ref{eq:alpha}) with $\alpha \ge 0$ ceases to exist: 
\begin{equation} \label{eq:gammaALLN}
	\gamma_{\rm AL} = 
	\left \{
	\begin{array}{ll}
		4/(2 - p) & \textrm{for~} p \le 1 \, , \\
		4 p &  \textrm{for~} p > 1 \, . 
	\end{array}
	\right .
\end{equation}
Hence for $p>1$ $\gamma_{\rm ergo}$ and $\gamma_{\rm AL}$ merge, the NEE disappears, and one has a discontinuous transition between the ergodic and the AL phases.

However, the expression found above for $\gamma_{\rm ergo}=4p$ does not give the correct result $\gamma_{\rm ergo} \to 1$ in the Gaussian RP limit $p \to 0$.
In fact, as we have already seen in the case of the L\'evy-RP ensemble (see Sec.~\ref{sec:PD}), the extreme value statistics argument used to determine $E_{\rm Th}$ as the maximum hybridization gap only applies if the tails of the off-diagonal elements are fat enough, i.e. $\mu<2$ for the L-RP case. If $\mu>2$, instead, Eq.~(\ref{eq:self-consistent}) underestimates the Thouless energy, which is alternatively given by the Fermi golden rule, $E_{\rm Th} \sim N \langle | {\cal H}_{ij} |^2 \rangle_W$. Similarly, in the LN-RP case for $p<1/2$ the width of the mini-bands is much larger than $N^{-\alpha}$ and is given by $E_{\rm Th} = \Gamma_{\rm av} \sim N^{1 - \gamma(1-p)}$. Requiring that the ergodic transition occurs when the Thouless energy becomes of the order of the total spectral bandwith one finally gets:
\begin{equation} \label{eq:gammaergoLN}
	\gamma_{\rm ergo} = 
	\left \{
	\begin{array}{ll}
		1/(1 - p) & \textrm{for~} p \le 1/2 \, , \\
		4 p &  \textrm{for~} p > 1/2 \, . 
	\end{array}
	\right .
\end{equation}
Eqs.~(\ref{eq:gammaergoLN}) and~(\ref{eq:gammaALLN}) are in perfect agreement with the results of Refs.~\cite{kravtsov1,khay}, although they have been obtained with a different approach.
However, the estimation of the Thouless energy given by the Fermi golden rule for $p<1/2$ is not expected to hold for $\gamma>2$. In fact for $\gamma>2$ the typical bandwidth $N [{\cal H}_{ij}]_{\rm typ}^2 \sim N^{1-\gamma}$ becomes smaller than the mean level spacing. In other words,  the typical value of the matrix elements $N^{-\gamma/2}$ is much smaller than the gap between neighboring levels and the system is essentially alike a sparse graph, where most of the matrix elements are effectively equal to $0$ in the thermodynamic limit. The Fermi golden rule is not expected to provide the correct estimation of the effective bandwidth in this regime (Eqs.~\eqref{eq:AL} and~\eqref{eq:FGR} only work for dense matrices)~\cite{remark} and one should thus switch back to $E_{\rm Th} \sim N^{-\alpha}$, with $\alpha$ given by Eq.~\eqref{eq:alpha}. (Note that one does not have this issue in the L-RP ensemble studied in the main text since in this case the typical bandwidth always becomes equal to the mean level spacing at the AL transition.) 

Hence, imposing that $E_{\rm Th} \sim N^{D_1-1}$ one obtains an estimation of the fractal dimension $D_1$ for the LN-RP ensemble:
\begin{equation} \label{eq:D1LN}
	D_1 = \left \{
	\begin{array}{ll}
		2 - \gamma(1 - p) & \textrm{for~} \gamma \le 2 \textrm{~and~} p \le 1/2 \, , \\
		1 - \alpha(\gamma,p) &  \textrm{for~} \gamma > 2 \textrm{~or~} p > 1/2 \, . 
	\end{array}
	\right .
\end{equation}
The equation above predicts that the fractal dimension $D_1$ is equal to one at the ergodic transition and exhibits a discontinuous jump at the AL transition, which is also in agreement with the findings of Refs.~\cite{kravtsov1,khay}. We find, however, a different value of the jump:
\[
D_1(\gamma = \gamma_{\rm AL}) = \frac{p}{2 - p} \, ,
\]
which goes to zero in the RP limit $p \to 0$ and to $1$ at the triciritical point $p \to 1$, but is strictly smaller than the value predicted in Ref.~\cite{khay}.

	
	\begin{figure*}
		\hspace{-0.5cm} \includegraphics[width=0.99\textwidth]{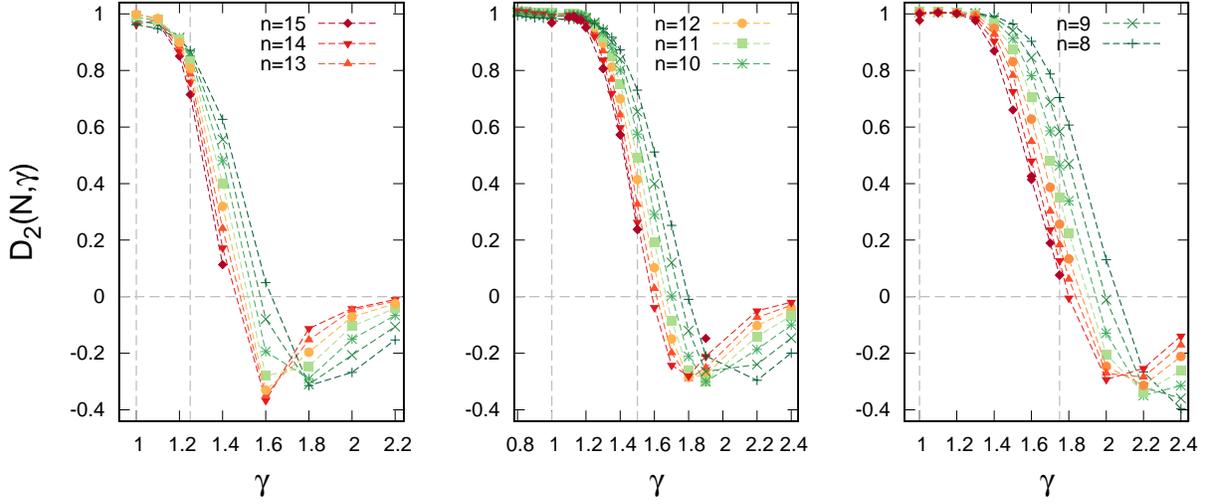}
		\vspace{-6cm}
		\caption{\label{fig:D2} Flowing $N$-dependent fractal dimensions $D_2 (N,\gamma)$ 
			for $\mu=1.25$ (left),  $\mu=1.5$ (middle),  $\mu=1.75$ (right),  and $N=2^n$ with $n=8,\ldots,15$, computed via Eq.~(\ref{eq:D2ed}) from the scaling  of the second moment of the wave-functions' amplitudes measured from exact diagonalizations.}
	\end{figure*}

\section{Fractal dimension $D_2$ and the finite-size scaling analysis of the moments of wave-functions' amplitudes} \label{app:D}

In this appendix we show few more numerical results on the fractal dimensions and on their finite-size scaling behavior.

\begin{figure*}
	\includegraphics[width=0.355\textwidth]{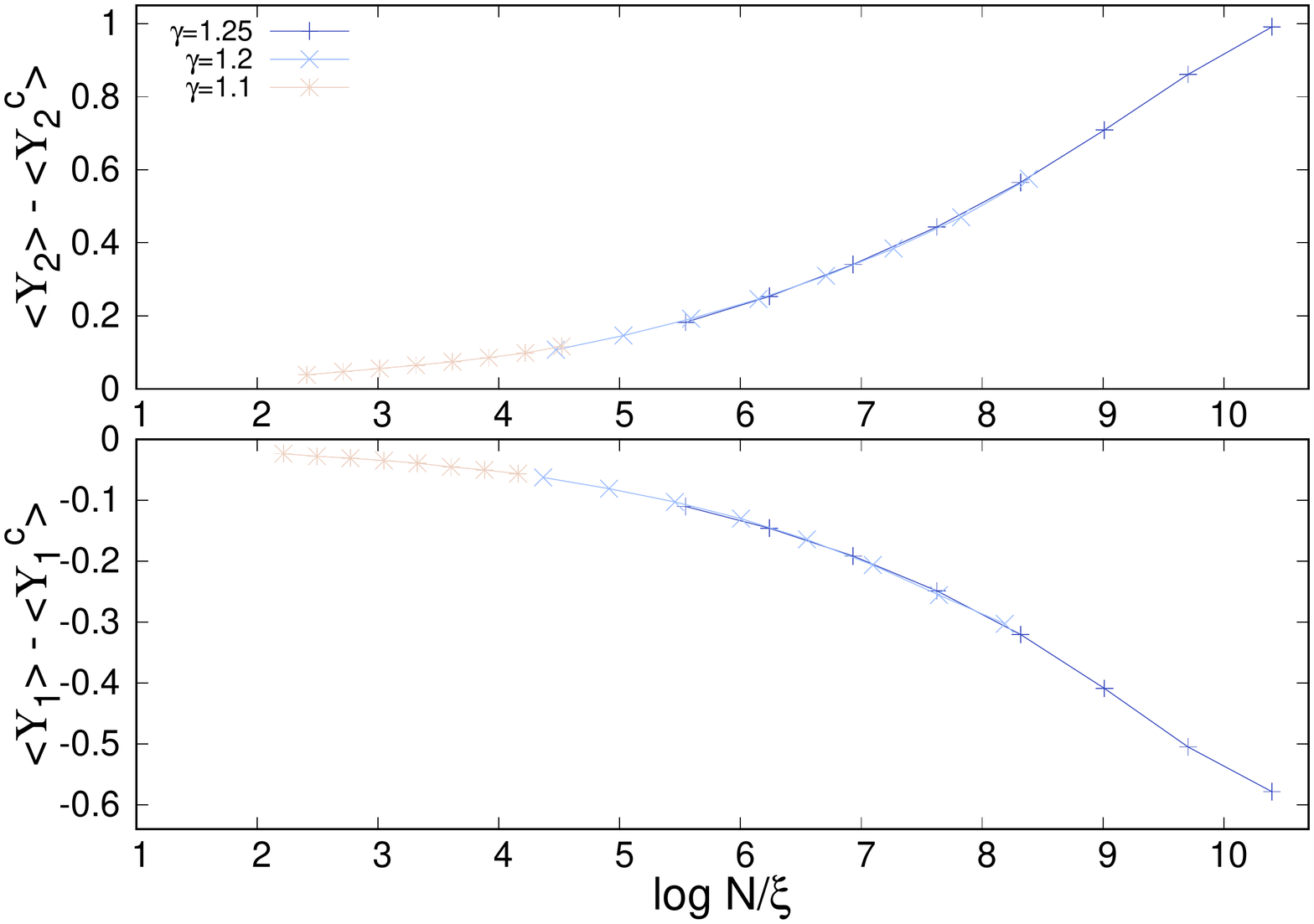}
	\hspace{-0.9cm}
	\includegraphics[width=0.355\textwidth]{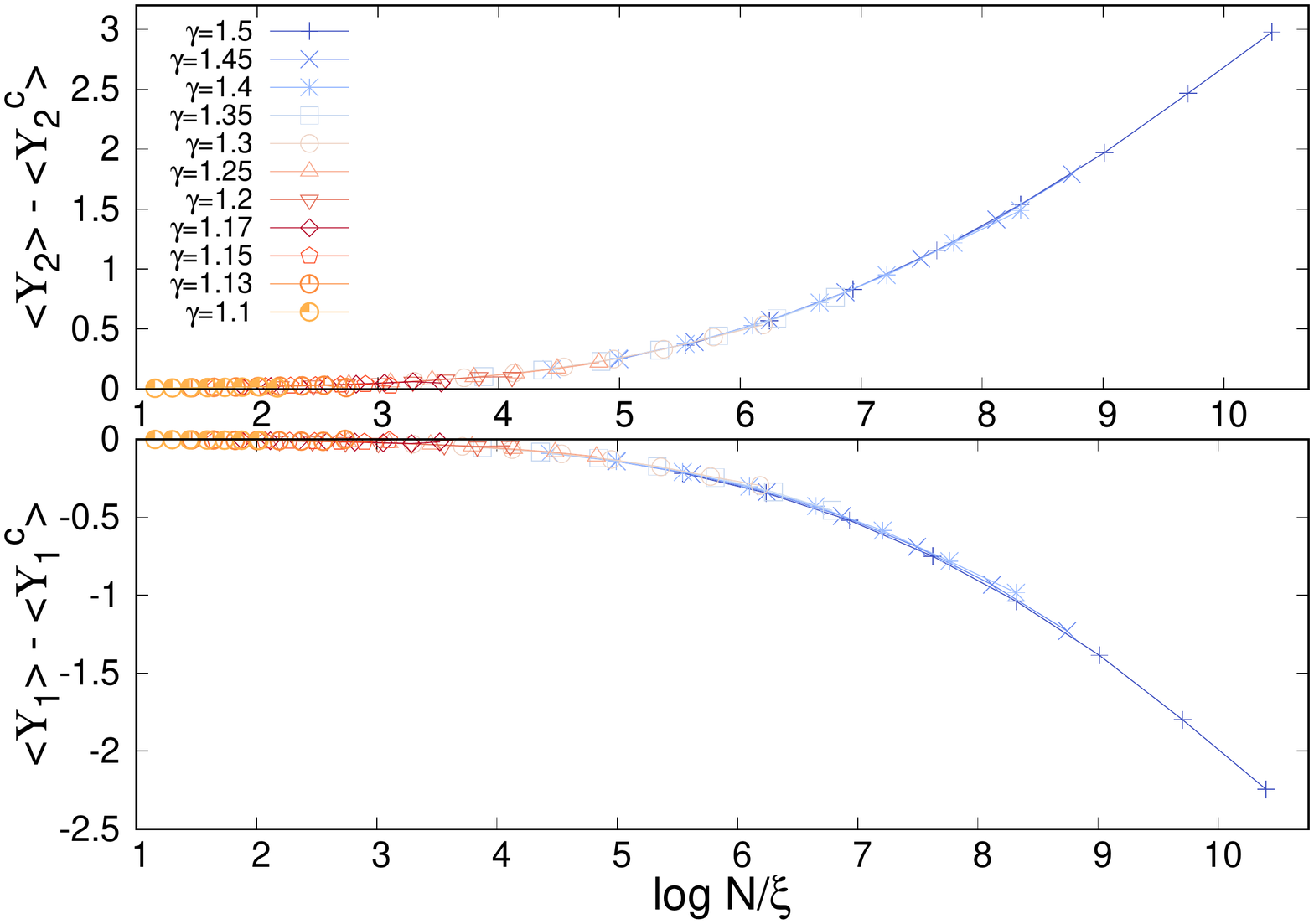}
	\hspace{-0.9cm}
	\includegraphics[width=0.355\textwidth]{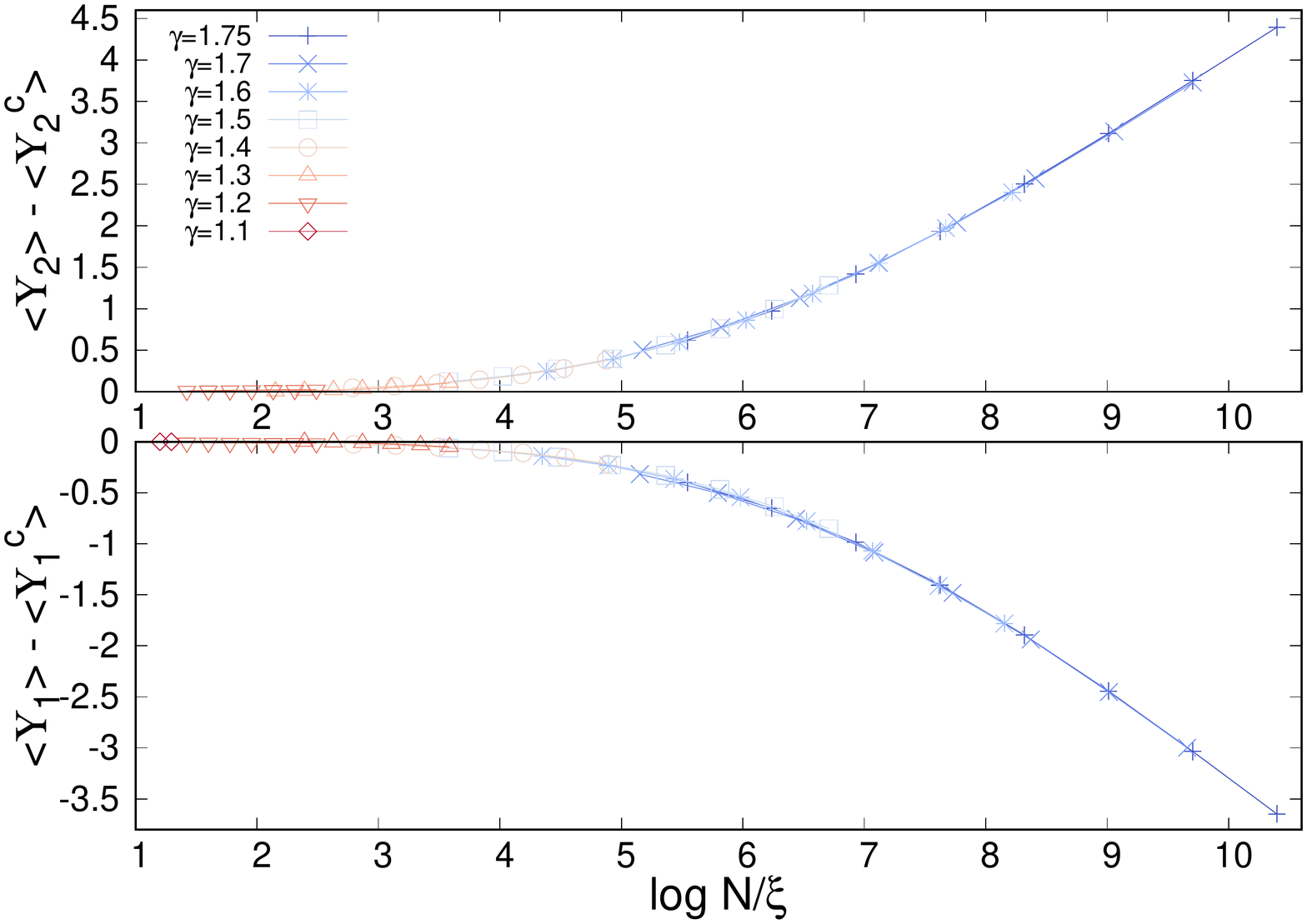}
	\caption{\label{fig:D} 
		Scaling curves for the first (bottom) and second (top) moments of the wave-functions amplitudes varying $\gamma \in (1,\mu)$ in the NEE phase for $\mu=1.25$ (left), $\mu=1.5$ (middle), and $\mu=1.75$ (right) according to Eq.~(\ref{eq:scaling}) and with $\xi$ given by Eq.~(\ref{eq:xi}). }
\end{figure*}

In Fig.~\ref{fig:D2} we plot the flowing fractal exponent $D_2 (N,\gamma)$  as a function of $\gamma$ for three values of $\mu \in (1,2)$ and several system sizes. 
$D_2 (N,\gamma)$ is estimated using Eq.~(\ref{eq:D2ed}) from the scaling with $N$ of the IPR, and its behavior is qualitatively similar to the one of  $D_1 (N,\gamma)$, shown in Fig.~\ref{fig:Df}.

Next we present an independent estimation of the value of the critical exponent $\nu_{\rm ergo}$ which describes the critical scaling  of the anomalous dimensions close to the transition point with the ergodic phase, $\gamma_{\rm ergo} = 1$.
This analysis is inspired by the one proposed in Ref.~\cite{mace} (see also Ref.~\cite{gabriel}) on the insulating side of  the MBL transition, where the fractal dimensions are decreasing functions of the disorder.
More precisely we posit that in the NEE phase, $1<\gamma<\mu$, 
the moments $\langle \Upsilon_q \rangle$ behave as:
\begin{equation} \label{eq:scaling}
	\begin{aligned}
		\langle \Upsilon_1 (N,\gamma) \rangle -  \langle \Upsilon_1 (N,\gamma=1)\rangle &= -D_{1,c} \frac{\log N}{\xi (\gamma)} \, ,\\
		\langle \Upsilon_q  (N,\gamma) \rangle -  \langle \Upsilon_q (N,\gamma=1)\rangle &= (q - 1) D_{q,c} \frac{\log N}{\xi (\gamma)} \, ,
	\end{aligned}
\end{equation}
with $D_{q,c}$ 
being the fractal dimensions at the transition point. 
The length scale $\xi$ depends on the distance to the critical point $\gamma_{\rm ergo} = 1$ and lies in the range $(1,+\infty)$,  which guarantees the fractal dimensions  to remain positive. The scaling ansatz above implies that in the limit $\log N \gg \xi$ the leading terms follows $\langle \Upsilon_1 \rangle \sim D_{1,c}( 1 - 1 / \xi (\gamma)) \log N$ and $\langle \Upsilon_q \rangle \sim - (q-1) D_{q,c}( 1 - 1 / \xi (\gamma)) \log N$, while in the opposite limit,  $\log N \ll \xi$, one retrieves the critical scaling. In order for Eqs.~(\ref{eq:D1}) and~(\ref{eq:Dinf}) to be satisfied one then should have that in the NEE phase 
\begin{equation} \label{eq:xi}
	\xi = \frac{\mu-1}{\gamma - 1} \, .
\end{equation}
Note that $\xi(\gamma=\mu) = 1$, in such a way that $D_{q} \to 0$ for $N \to \infty$ at the Anderson localization. 
As shown in Fig.~\ref{fig:D}, a very good data collapse is obtained in the intermediate phase $\gamma \in (1, \mu)$  and for all values of $\mu$  when the first and second moments of the wave-functions amplitudes are plotted as a function of the scaling variable $\log N / \xi$,  where  $\xi(\gamma)$ is chosen as in Eq.~(\ref{eq:xi}), confirming that the critical exponent $\nu_{\rm ergo}$ is equal to one at the ergodic transition independently of $\mu$. Notice that there is no adjustable parameter in this procedure. 

Note that the finite-size scaling of Fig.~\ref{fig:D} with $\xi$ given by Eq.~\eqref{eq:xi} automatically implies that for the L-RP model in the thermodynamic limit, $\log N \gg \xi$, the ratio $D_q/D_1$ is equal to $D_{q,c}/D_{1,c}$ independently of $\gamma$. Hence, from Eq.~\eqref{eq:D1} one thus has that for $1 < \mu <2$ and $1 < \gamma < \mu$ the fractal dimensions $D_q$ are also straight lines vanishing at $\gamma=\mu$ with $q$-dependent slopes: $D_q(\gamma) = D_{q,c} (\mu - \gamma)/(\mu - 1)$. This is fully consistent with our prediction~\eqref{eq:Dinf}, with $D_{\infty,c}= 2(\mu-1)/\mu$. If $\log N \ll \xi$, however, one does not observe that the ratio $D_q/D_1$ is constant (see e.g. Fig.~\ref{fig:DfinfD6}) due to the fact that the scaling functions for different values of $q$ are different, producing different $q$-dependent finite size effects.

\begin{figure}
	\includegraphics[width=0.49\textwidth]{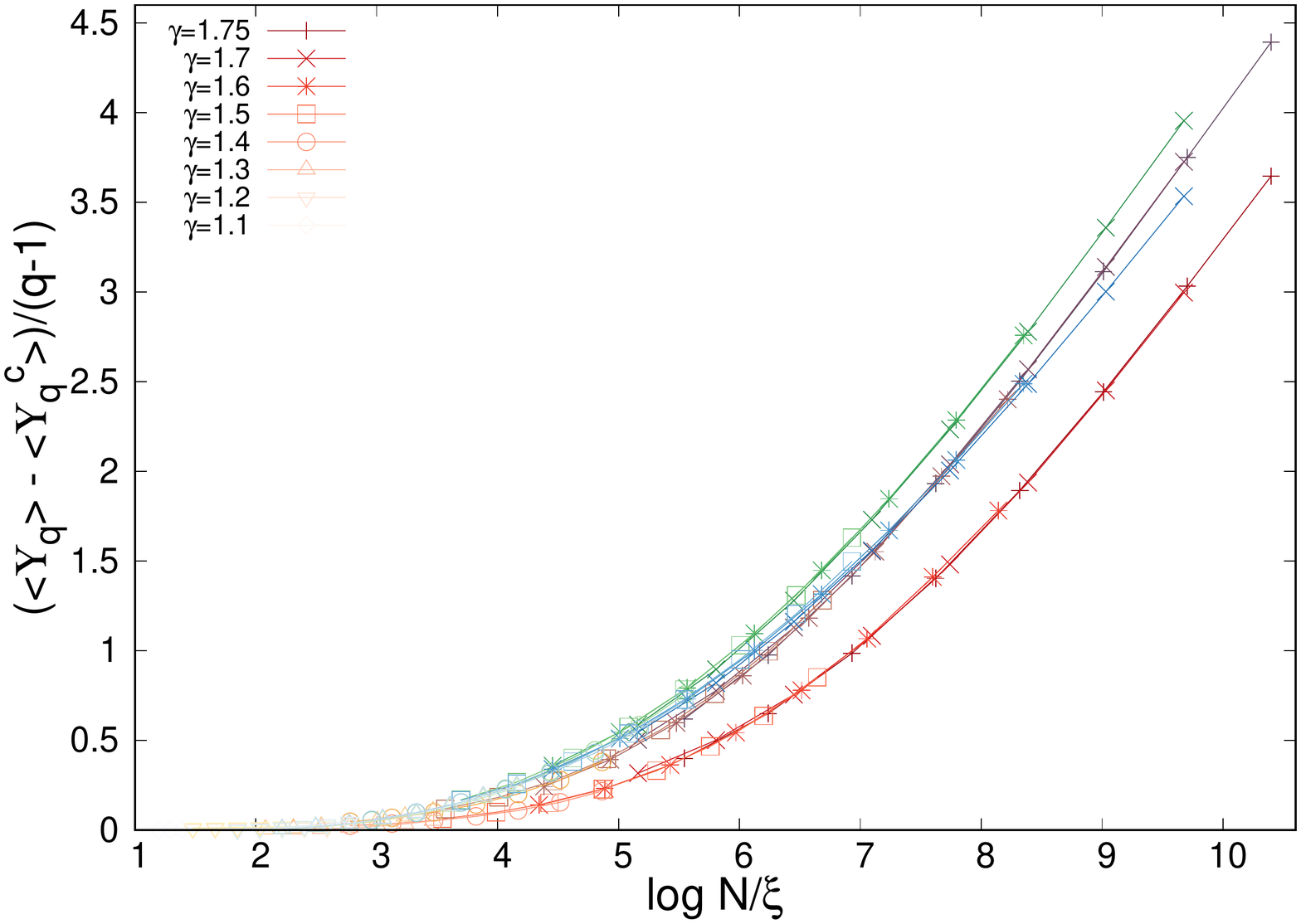}
	\caption{\label{fig:D126inf} 
		Scaling curves for the  $q$-th moments of the wave-functions amplitudes (divided by $q-1$) varying $\gamma \in (1,\mu)$ in the NEE phase for $\mu=1.75$ according to Eq.~(\ref{eq:scaling}) and with $\xi$ given by Eq.~(\ref{eq:xi}), for $q=1$ (red), $q=2$ (brown), $q=6$ (green), and $q \to \infty$ (blue). For $q=1$ we have changed the signs of $\Upsilon_1$ in order to have all the data on the positive side of the $y$-axis.}
\end{figure}

This is confirmed by Fig.~\ref{fig:D126inf}, where we plot the same finite-size scaling analysis for different moments of the wave-functions' amplitudes and  for $\mu=1.75$, showing that a very good collapse is found for all values of $q$ in terms of the scaling variable $\log N / \xi$. The scaling functions depend on $q$ and  the fact that the scaling function for $q \to \infty$ approaches a straight line for $\log N / \xi \gg 1$ with  a smaller slope  compared to the scaling function for $q=1$ indicates that $D_{\infty,c}<D_{1,c}$, in agreement with the fact that $D_{\infty}$ has a discontinuous jump at the ergodic transition.

\section{Perturbation theory and the spectrum of fractal dimensions in the AL phase} \label{sec:first}
In this appendix we discuss the standard perturbation theory for 
the amplitudes $w_{ij} = | \psi_i (j)|^2$, focusing in particular on its domain of convergence.
This calculation allows one to obtain the full spectrum of fractal dimensions in the AL phase where the preturbative expansion converges absolutely.
We focus on the region $\mu<2$, since the case $\mu>2$ is equivalent to the RP ensemble treated in Ref.~\cite{kravtsov}.

The first order perturbation theory gives:
\[
| \psi_i \rangle = | i \rangle + \sum_{j(\neq i)} \frac{{\cal L}_{ij}}{\epsilon_i - \epsilon_j} | j \rangle \, .
\]
The amplitude on the sites $j \neq i$ is therefore given by:
\begin{equation} \label{eq:first}
w_{ij} = \frac{{\cal L}_{ij}^2}{(\epsilon_i - \epsilon_j)^2} \equiv R_{ij}^2 \, ,
\end{equation}
where $R_{ij} = {\cal L}_{ij} / \delta_{ij}$ are the hybridization ratios between the levels $i$ and $j$ (with $\delta_{ij} \equiv \epsilon_i - \epsilon_j$).
The typical value of $R_{ij}$ is 
\[
R_{ij}^{\rm typ} = \frac{{\cal L}_{ij}^{\rm typ}}{\delta_{ij}^{\rm typ}} \sim \frac{N^{-\gamma/\mu}}{W} 
\to 0 \, .
\]
However, large hybridization ratios can be obtained from the biggest off-diagonal coupling:
\[
R_{ij}^{\rm max} = \frac{{\cal L}_{ij}^{\rm max}}{\delta_{ij}^{\rm typ}} \sim \frac{N^{(1-\gamma)/\mu}}{W} 
\, ,
\]
which  vanish in the thermodynamic limit for $\gamma > 1$. 
Finally, large hybridization ratios can be obtained from small energy difference (i.e., consecutive levels):
\[
R_{ij}^{\rm next} = \frac{{\cal L}_{ij}^{\rm typ}}{\delta_{ij}^{\rm min}} \sim \frac{N^{-\gamma/\mu}}{W N^{-1}} \sim N^{1-\gamma/\mu} \, ,
\]
which vanish in the thermodynamic limit for $\gamma > \mu$.
As a result, the perturbative series converge absolutely for $\mu>1$ and $\gamma>\mu$, and for  $\mu<1$ and $\gamma>1$, as indicated in Fig.~\ref{fig:PD}, 
since the average  off-diagonal matrix elements 
times the coordination number $N$ is much smaller than the typical difference of the diagonal matrix elements [recovering the Mott's criterion~(\ref{eq:AL})].
For $\mu>1$ and $\gamma < \mu$ the convergence of the series might still occur  because of the random and independently
fluctuating signs of ${\cal L}_{ij}$ and $\delta_{ij}$, as in the RP model with Gaussian elements.
In order to analyze this possibility, we perform the calculation at second order, which gives:
\[
w_{ij} \approx \delta_{ij} + \frac{1}{\epsilon_i - \epsilon_j} \left[  {\cal L}_{ij} + \sum_{k \neq i} \frac{{\cal L}_{jk} {\cal L}_{ki}}{\epsilon_i - \epsilon_k}  + \ldots \right] \, .
\]
Since ${\cal L}_{jk} {\cal L}_{ki}$ 
is the product of two uncorrelated L\'evy-distributed random variables, we can  apply the generalized central limit theorem to characterize the probability distribution of the sum in the square brackets, which is again a L\'evy distributed random variable with exponent $1+\mu$ and typical value 
 of order $N^{1 - 2 \gamma / \mu}$.
 For $\mu>1$ the typical value dominates the average, 
in agreement with the arguments given in Sec.~\ref{sec:estimation}.
(For $\mu<1$ instead this term is of order $N \langle {\cal L}_{ij} \rangle^2_W \sim N^{1 - 2 \gamma}$).
This term is much bigger than the typical value of the first term, $N^{-\gamma / \mu}$, if $\gamma < \mu$. 
 By applying the same kind of reasoning it is straightforward to generalize this calculation to the higher order terms of the perturbative expansion, showing that the 
 the terms of order $n$ in the square brackets above are L\'evy distributed random variables with exponent $1+ \mu$ and typical value scaling as $N^{n-1} (\langle {\cal L}_{ij} \rangle_W)^n \sim N^{n(1 - \gamma/\mu) - 1}$ for $\mu>1$.  
 (For $\mu<1$ instead this term scales as $N^{n(1-\gamma)-1}$). Note that in the RP limit, $\mu > 2$, the higher order terms are instead negligible in the thermodynamic limit thanks to the random signs of the matrix elements which ensures that $\langle {\cal L}_{jk} {\cal L}_{ki} \rangle = 0$, implying that the first order computation gives the correct results at large $N$ in the Gaussian case.
 
 It is also instructive to analyze the pertrubative expansions for the eigenvalues, which reads:
 \[
 \lambda_i \approx \epsilon_i + \sum_{j \neq i} \frac{{\cal L}_{ij}^2}{\epsilon_i - \epsilon_j} + \sum_{j \neq i}  \sum_{k \neq  i}  \frac{{\cal L}_{ij} {\cal L}_{jk} {\cal L}_{ki}}{(\epsilon_i - \epsilon_j)(\epsilon_i - \epsilon_k)} \ldots \, .
 \]
The second term of the r.h.s. of the expression above is a L\'evy distributed random variable with power law exponent $1 + \mu/2$ and typical value of order $N^{1 - 2 \gamma / \mu}$. By applying the arguments of Sec.~\ref{sec:estimation} one obtains that, due to the power-law tails of the distribution, the average amount of energy that the levels move at second order is $N \langle {\cal L}_{ij}^2 \rangle_W = N^{1-\gamma}$ (where the average is cut at the spectral bandwidth).
Note that, differently from the RP model, $N \langle {\cal L}_{ij}^2 \rangle_W \neq N \langle {\cal L}_{ij} \rangle_W^2$, implying that the typical and the average bandwidth do not coincide.
The amplitude of higher order terms can be evaluated as above. At order $n$ one has $N^{n-1} (\langle {\cal L}_{ij} \rangle_W)^n \sim N^{n(1 - \gamma/\mu) - 1}$ for $\mu>1$.

\subsection{The NEE phase}

Hence, due to the fat-tails distribution of the off-diagonal matrix elements, differently from its Gaussian RP counterpart, in the L-RP ensemble the higher order terms of the perturbative series cannot be neglected in the NEE regime $1<\mu<2$ and $\gamma < \mu$. In the following we show that in fact keeping only the first-order term leads to a wrong result for the probability distribution of the wave-functions' amplitudes and the anomalous dimensions.

At first order, Eq.~(\ref{eq:first}),  the $w_{ij}$'s are given by the product of two power-law distributed random variables:
$x_{ij} = {\cal L}_{ij}^2$ have a power-law tail with an exponent $1 + \mu/2$ and typical value $N^{-2 \gamma / \mu}$:
\[
P(x_{ij}) = \frac{\mu}{2 N^\gamma x_{ij}^{1+\mu/2}} \, \theta (x_{ij} > N^{-2 \gamma/\mu}) \, ,
\]
and $y_{ij} = \delta_{ij}^{-2}$ have typical value of $O(W)$ and power-law tails with an exponent $3/2$:
\[
P(y_{ij}) = \frac{e^{-1/(4 W^2 y_{ij})}}{\sqrt{4 \pi W^2}} y_{ij}^{-3/2} \, .
\]

For $\mu > 1$ the amplitudes $w_{ij}$ are power-law distributed with an exponent $3/2$ and typical value $w_{ij}^{\rm typ} = x_{ij}^{\rm typ} y_{ij}^{\rm typ} = N^{-2 \gamma / \mu}$. Without loss of generality the distribution can be written as:
\[
P(w_{ij}) = \frac{1}{w_{ij}^{\rm typ}} P_{\rm reg} \left( \frac{w_{ij}}{w_{ij}^{\rm typ}} \right ) + c \, \frac{\theta (w_{ij} >  w_{ij}^{\rm typ})}{N^{\gamma / \mu} \, w_{ij}^{3/2}}  \, .
\]
The normalization of wave-functions imposes that $\langle w_{ij} \rangle = N^{-1}$, which implies an  upper cut-off $w_{ij}^{\rm max}$ to the singular part of the distribution above:
\[
N \langle w_{ij} \rangle \sim N^{1 - 2 \gamma/\mu} + \frac{c}{N^{\gamma/ \mu - 1}} \int_{ N^{-2 \gamma / \mu}}^{w_{ij}^{\rm max}} \! \! w_{ij}^{-1/2} {\rm d} w_{ij} = 1 \, ,
\]
which yields:
\[
w_{ij}^{\rm max} \sim N^{2 (\gamma / \mu - 1)} \, .
\]
A caution, however, should be taken, since the amplitudes $w_{ij}$
on any lattice site cannot exceed $1$. Hence, the above estimation of $w_{ij}^{\rm max}$ is only correct if $2 (\gamma / \mu - 1) < 0$, i.e., $\gamma < \mu$, while for $\gamma>\mu$ we have that $w_{ij}^{\rm max} =1$.
In order to compensate for the deficiency of normalization of $\langle w \rangle$ in the latter case one has to assume a singular
part of the distribution:
\[
\hat{P}(w_{ij}) = P(w_{ij})+A \delta (w_{ij}-1) \, .
\] 
One can see that for
$\gamma>\mu$  the average  $\langle w \rangle$  is dominated by the singular term, and $A = N^{-1}$. This corresponds to the strongly localized wave
functions $\psi_i (j)$ on the site $i$.

At this point one can easily compute the spectrum of fractal dimensions $f(\alpha)$, describing the number of amplitudes scaling as $N^{-\alpha}$. For $\gamma < \mu$ we have:
\[
N^{f(\alpha)} = \frac{c}{N^{\gamma/ \mu - 1}}  \int_{N^{-\alpha}}^{ N^{2 (\gamma / \mu - 1)}} \! \! \! w_{ij}^{-3/2} {\rm d} w_{ij} \sim N^{\alpha/2 - \gamma / \mu + 1} \, ,
\]
for $2(1 - \gamma / \mu) = \alpha_{\rm min} < \alpha < \alpha_{\rm max} = 2 \gamma / \mu$, which gives:
\[
f(\alpha) = \frac{\alpha}{2} - \frac{\gamma} { \mu } + 1 \, \, \, \, \, \, \, \,  (\alpha_{\rm min} < \alpha < \alpha_{\rm max} ) \, .
\]
In the localized region $\gamma > \mu$, $\alpha_{\rm min} = 0$. At the AL transition point the function $f(\alpha) = \alpha/2$  has the same triangular shape as the Anderson model on the Bethe lattice at the localization transition.
In the region of the extended non-ergodic states, $1 < \gamma < \mu$, $\alpha_{\rm min} > 0$.

Alternatively, one can compute directly the moments $N \langle | \psi_i (j) |^{2 q} \rangle \sim N^{- \tau_q}$:
\[
\langle w^q \rangle \sim N^{- 2  q \gamma/\mu} + \frac{c}{N^{\gamma/ \mu}} \int_{ N^{-2 \gamma / \mu}}^{ N^{2 (\gamma / \mu - 1)}} \! \! \! w_{ij}^{q - 3/2} {\rm d} w_{ij} \, .
\]
For $q<1/2$ the moments $\langle w^q \rangle$ are dominated by the typical values and $\tau_q = 2 q \gamma / \mu - 1$, while for $q>1/2$ the moments are dominated by the upper cut-off and
$\tau_q = 2 (q - 1) (1 - \gamma / \mu)$. 
One can thus compute the fractal dimensions $D_q = \tau_q/ (q-1)$:
\begin{equation} \label{eq:Dfirst}
D_q = \left \{
\begin{array}{ll}
2 (1 - \gamma/ \mu) & \textrm{for~} q>1/2 \, ,\\
\frac{1 - 2 q \gamma / \mu}{1 - q} & \textrm{for~} q<1/2 \, .
\end{array}
\right .
\end{equation}
Thus the first-order expression does not coincide with the one found fin the main text, Eq.~(\ref{eq:D1}), and 
corresponds to non-ergodic extended states for $1 < \gamma < \mu$ that occupy a fraction $N^{1 - 2 \gamma / \mu}$ of sites only.
This is due to the fact that the first-order computation neglects the effect of anomalously large matrix elements that can hybridize sites at energy distance much larger than $N [{\cal H}_{ij}^2]_{\rm typ}$.
The main difference is that the first-order calculation predicts 
that $D_q$ approaches a value smaller than one, $D_c = 2 (1 - 1 / \mu)$, at the ergodic transition $\gamma_{\rm ergo} = 1$ for all $q>1/2$, where
$D_c = 2 (1 - 1 / \mu)$, with a discontinuous jump at the transition. This also predicts that, as for the RP model, the fractal dimensions are degenerate.

\subsection{The localized phase}

In the localized phase, $\gamma > \mu$, the pertrubative expansion does converge. Repeating the calculation above one obtains the spectrum of fractal dimensions as~\cite{monthus-LRP}:
\begin{equation} \label{eq:Dloc}
D_q = \left \{
\begin{array}{ll}
0 & \textrm{for~} q> \mu/(2 \gamma) \, ,\\
\frac{1 - 2 q \gamma / \mu}{1 - q} & \textrm{for~} q< \mu/ ( 2 \gamma) \, .
\end{array}
\right .
\end{equation}
A similar behavior is also found for the Anderson model on the Bethe lattice.

For $\mu < 1$  and $\gamma>1$,  going back to Eq.~(\ref{eq:first}) one has that at first order in perturbation theory the amplitudes $w_{ij}$ are power-law distributed with an exponent $1 + \mu/2$ and typical value $w_{ij}^{\rm typ} = x_{ij}^{\rm typ} y_{ij}^{\rm typ} = N^{-2 \gamma / \mu}$:
\[
P(w_{ij}) = \frac{1}{w_{ij}^{\rm typ}} P_{\rm reg} \left( \frac{w_{ij}}{w_{ij}^{\rm typ}} \right ) + c \, \frac{\theta (w_{ij} >  w_{ij}^{\rm typ})}{N^{\gamma } \, w_{ij}^{1 + \mu/2}}  \, .
\]
The computation for the moments of the wave-functions's amplitudes thus yields that the spectrum of fractal dimensions is the one given by Eq.~(\ref{eq:Dloc}).




\begin{thebibliography}{99}

\bibitem{wegner} F. Wegner, Z. Phys. B {\bf 36}, 209 (1980).

\bibitem{rodriguez} A. Rodriguez, L.J. Vasquez, K. Slevin, and R.A. Romer, Phys.
Rev. B {\bf 84}, 134209 (2011).

\bibitem{BAA} D.M. Basko, I.L. Aleiner, and B.L. Altshuler,
Ann. Phys. {\bf 321}, 1126 (2006).

\bibitem{Gornyi} I. V. Gornyi, A. D. Mirlin, and D. G. Polyakov, Phys. Rev. Lett. {\bf 95}, 206603 (2005).

\bibitem{reviewMBL1} E. Altman and R. Vosk, Annu. Rev. Condens. Matter Phys. {\bf 6}, 383 (2015).

\bibitem{reviewMBL2}
R. Nandkishore and D. A. Huse, Annu. Rev. Condens. Matter Phys. {\bf 6}, 15 (2015).

\bibitem{reviewMBL3} D. A. Abanin and Z. Papi\'c,  Annalen  der  Physik {\bf 529}, 1700169 (2017).

\bibitem{reviewMBL4}
F. Alet and N. Laflorencie, Comptes Rendus Physique {\bf 19}, 498 (2018).

\bibitem{reviewMBL5}  D.  A.  Abanin,  E.  Altman,  I.  Bloch, and  M. Serbyn, Rev. Mod. Phys. {\bf 91}, 021001 (2019).

\bibitem{alet} D. J. Luitz, N. Laflorencie, and F. Alet, Phys. Rev. B {\bf 93}, 060201(R) (2016).

\bibitem{mace} N. Mac\'e, F. Alet, and N. Laflorencie, Phys. Rev. Lett. {\bf 123}, 180601 (2019).

\bibitem{laflorencie} F. Pietracaprina and N Laflorencie, {\tt arXiv:1906.05709}

\bibitem{war} G. De Tomasi, I. M. Khaymovich, F. Pollmann, and S. Warzel, {\tt arXiv:2011.03048}

\bibitem{resonances1} I. V. Gornyi, A. D. Mirlin, D. G. Polyakov, and A. L. Burin, Annalen der Physik {\bf 529}, 1600360 (2017);

\bibitem{resonances2} K. S. Tikhonov and A. D. Mirlin, Phys. Rev. B {\bf 97}, 214205 (2018).

\bibitem{tarzia} M. Tarzia, Phys. Rev. B {\bf 102}, 014208 (2020).

\bibitem{subdiff1} D. J. Luitz and Y. Bar Lev, Ann. Phys. 1600350 (2017).

\bibitem{subdiff2} K. Agarwal, E. Altman, E. Demler, S. Gopalakrishnan, D. A. Huse, and M. Knap, Ann. Phys., 1600326 (2017).

\bibitem{dinamica} G. Biroli and M. Tarzia, Phys. Rev. B {\bf 96}, 201114(R) (2017).

\bibitem{levitov} B. L. Altshuler, Y. Gefen, A. Kamenev, L. S. Levitov,
Phys. Rev. Lett. {\bf 78}, 2803 (1997).

\bibitem{qrem1} L. Faoro, M. V. Feigel’man, and L. Ioffe, Annals of Physics {\bf 409}, 167916 (2019).

\bibitem{qrem2} C. L. Baldwin and C. R. Laumann, Phys. Rev. B {\bf 97}, 224201 (2018).

\bibitem{qrem3} V. Smelyanskiy, K. Kechedzhi, S. Boixo, H. Neven, and B. Altshuler, {\tt arXiv:1907.01609}

\bibitem{qrem4} T. Parolini and G. Mossi, {\tt arXiv:2007.00315}

\bibitem{qrem5} G. Biroli, D. Facoetti, M. Schir\'o, M. Tarzia, and P. Vivo, {\tt arXiv:2009.09817}

\bibitem{boixo} K. Kechedzhi, V. N. Smelyanskiy, J. R McClean, V. S Denchev, M. Mohseni, S. V. Isakov, S. Boixo, B. L. Altshuler, and H. Neven, {\tt arXiv:1807.04792}

\bibitem{jj} M. Pino, L. B. Ioffe, and B. L. Altshuler,
PNAS, {\bf 113}, 536 (2016);
M. Pino, V. E. Kravtsov, B. L. Altshuler, and L. B. Ioffe,
Phys. Rev. B, {\bf 96} 214205, (2017).

\bibitem{syk} T. Micklitz, F. Monteiro, and A. Altland,
Phys. Rev. Lett., {\bf 123} 125701 (2019);
F. Monteiro, T. Micklitz, M. Tezuka, and A. Altland,
{\tt arXiv:2005.12809}

\bibitem{rmt1} M. L. Metha, {\it Random Matrices}, 3rd ed. (Elsevier, Amsterdam, 2004).

\bibitem{rmt2} G. Akemann, J. Baik, and P. Di Francesco (eds.), {\it The Oxford Handbook of Random Matrix Theory} (Oxford University Press, Oxford, 2011).

\bibitem{kravtsov} V. E. Kravtsov, I. M . Khaymovich, E. Cuevas, M. Amini,
New Journal of Physics {\bf 17}, 122002 (2015).

\bibitem{RP} N. Rosenzweig and C. E. Porter, Phys. Rev. B {\bf 120}, 1698 (1960).

\bibitem{warzel} P. von Soosten and S. Warzel, Letters in Mathematical Physics 1, (2018). 

\bibitem{facoetti} D. Facoetti, P. Vivo, and G. Biroli, EPL {\bf 115}, 47003 (2016).

\bibitem{bogomolny} E. Bogomolny and M. Sieber, Phys. Rev. E {\bf 98}, 042116 (2018).

\bibitem{bera} G. de Tomasi, M. Amini, S. Bera, I. M. Khaymovich, and V. E. Kravtsov. SciPost Phys. {\bf 6}, 014 (2019). 

\bibitem{pino} M. Pino, J. Tabanera, and P. Serna, {\bf 52}, 475101 (2019).

\bibitem{shapiro1} A. Altland, M. Janssen, and B. Shapiro,
Phys. Rev. E {\bf 56}, 1471 (1997).

\bibitem{shapiro2} H. Kunz and B. Shapiro,
Phys. Rev. E {\bf 58}, 400 (1998).

\bibitem{roy} S. Roy and D. Logan, Phys. Rev. B {\bf 101}, 134202 (2020).

\bibitem{kravtsov1} V. E. Kravtsov, I. M. Khaymovich, B. L. Altshuler, L. B. Ioffe, {\tt arXiv:2002.02979}

\bibitem{khay} I. M. Khaymovich, V. E. Kravtsov, B. L. Altshuler, and L. B. Ioffe, {\tt arXiv:2006.04827}

\bibitem{monthus-LRP} C. Monthus, 
J. Phys. A: Mathematical and Theoretical {\bf 50}, 295101 (2017).

\bibitem{JP} P. Cizeau and J.-P. Bouchaud, Phys. Rev. E {\bf 50}, 1810
(1994). 

\bibitem{footnote3} For concreteness one can take $p(\epsilon) = e^{-\epsilon^2/(2 W^2)}/\sqrt{2 \pi W^2}$, or a box distribution in the interval $[-W/2,W/2]$, with $W$ of order $1$.

\bibitem{noi} E. Tarquini, G. Biroi, and M. Tarzia, Phys. Rev. Lett. {\bf 116}, 010601 (2016).

\bibitem{benarous} G. Ben Arous and A. Guionnet, Comm. in Math. Phys.
{\bf 278}, 715 (2008).

\bibitem{burda} Z. Burda, J. Jurkiewicz, M.A. Nowak, G. Papp, and I.
Zahed, Phys. Rev. E {\bf 75}, 051126 (2007).

\bibitem{metz} F.L. Metz, I. Neri, and D. Boll\`e, Phys. Rev. E {\bf 82}, 031135
(2010); J. Stat. Mech. (2010) P01010.

\bibitem{monthus} C. Monthus, JSTAT 093304 (2016).

\bibitem{edges} G. Biroli, J.-P. Bouchaud, M. Potters, Europhys. Lett. {\bf 78}, 78 (2007).

\bibitem{auff} A. Auffinger, G. Ben Arous, and S. P\'ech\'e, Annales de
l’IHP - Probabilit\'es et Statistiques {\bf 45}, 589 (2009).

\bibitem{borde} C. Bordenave and A. Guionnet, Probab. Theory and Relat. Fields {\bf 157}, 885 (2013).

\bibitem{cauchy} S. Majumdar, G. Schehr, D. Villamaina, and P. Vivo, J. Phys. A {\bf 46}, 022001 (2013).

\bibitem{lopatto1}
A. Aggarwal, P. Lopatto, H.T. Yau, {\tt arXiv:1806.07363}

\bibitem{lopatto2}
A. Aggarwal, P. Lopatto, J. Marcinek, {\tt arXiv:2002.09355}

\bibitem{abou} R. Abou-Chacra, P. W. Anderson, and D. J. Thouless, J. Phys. C {\ bf 6}, 1734 (1973).

\bibitem{susy} Y.V. Fyodorov and A.D. Mirlin, J. Phys. A {\bf 24}, 2273 (1991); Phys. Rev. Lett. {\bf 67}, 2049 (1991); Y. V. Fyodorov, A. D. Mirlin, and H.-J. Sommers, Journal de Physique I {\bf 2}, 1571 (1992).

\bibitem{Bethe} G. Biroli and M. Tarzia, {\tt arXiv:1810.07545}

\bibitem{tikhonov} K. S. Tikhonov and A. D. Mirlin, Phys. Rev. B {\bf 99}, 024202 (2019).

\bibitem{footnote1} In fact for $\mu < 1$ the fractal dimension $D_1$ is expected to exhibit a discontinuous jump at the AL transition from $D_1 = 1$ for $\gamma \le 1$ to $D_1 = 0$ for $\gamma  > 1$ at small energy $|E| < E_{\rm loc} (\mu, \gamma) = E_{\rm loc} (\mu) N^{(1 - \gamma)/\mu}$. At high energy, $|E| > E_{\rm loc} (\mu, \gamma) = E_{\rm loc} (\mu) N^{(1 - \gamma)/\mu}$, instead, one has a discontinuous transition at $\gamma=1$ between two different AL phases and $D_1$ should display a discontinuous jump from $D_1 (\mu, E)$ for $\gamma \le 1$ to $D_1 = 0$ for $\gamma>1$.

\bibitem{nosov} P. A. Nosov, I. M. Khaymovich, and V. E. Kravtsov, Phys. Rev. B {\bf 99}, 104203 (2019).

\bibitem{kramir1} V. E. Kravtsov and A. D. Mirlin , JETP Lett. {\bf 60}, 656 (1994).

\bibitem{kramir2} Y. V. Fyodorov and A. D. Mirlin,
Phys. Rev. B {\bf 51}, 13403 (1995).

\bibitem{prefactor} The prefactor $N^{D_1}$ is such that for $\eta=N^{-1}$ one has ${\rm Im}s {\cal G}_{\rm typ} = N^{D_1 - 1}$ matching the regime $N^{-1} \ll \eta \ll E_{\rm Th}$ of Eq.~\eqref{eq:Th}.

\bibitem{huse} V. Oganesyan, D. Huse, Phys. Rev. B {\bf 75}, 155111 (2007).

\bibitem{gabriel} I. Garcia-Mata, O. Giraud, B. Georgeot, J. Martin, R. Dubertrand, and G. Lemarié,
Phys. Rev. Lett. {\bf 118}, 166801 (2017).

\bibitem{altshulerK2} B. L. Altshuler and B. I. Shklovskii, Zh. Eksp. Teor. Fiz.
{\bf 91}, 220 (1986); B. L. Altshuler and B. I. Shklovskii, Sov.
Phys. JETP {\bf 64}, 127 (1986).

\bibitem{chalker}  J. T. Chalker, Physica A {\bf 167}, 253 (1990); 
J. T. Chalker and G. J. Daniell, Phys. Rev. Lett. {\bf 61}, 593 (1988).

\bibitem{kravK2} E. Cuevas and V. E. Kravtsov, Phys. Rev. B {\bf 76}, 235119 (2007).

\bibitem{mirlin} A. D. Mirlin, Physics Reports {\bf 326}, 259 (2000).

\bibitem{serbyn} M. Serbyn, Z. Papic', and D. A. Abanin,  Phys. Rev. B {\bf 96}, 104201 (2017).

\bibitem{polk} D. Sels and A. Polkovnikov, {\tt arXiv:2009.04501}

\bibitem{tikhK2} K. S. Tikhonov, A. D. Mirlin, {\tt arXiv:2009.09685}

\bibitem{footnote2} In the calculation of $K_2 (\omega)$ with the cavity approach, the imaginary regulator $\eta$ is set equal to $1/N$.

\bibitem{remark} Take for instance L-RP matrices with $\mu<1$ and $\gamma \le 1$, i.e. typical value $[{\cal H}_{ij}]_{\rm typ} \sim N^{-\gamma / \mu}$ much smaller than the mean level spacing $\Delta \sim N^{(1 - \mu - \gamma)/\mu}$. In this case the Fermi golden rule would suggest that the system is ergodic, since $N \langle | {\cal H}_{ij} |^2 \rangle_B \sim N^{2(1 - \gamma)/\mu}$ diverges, while one knows from~\cite{noi} that a mobility edge appears at energy $E_{\rm loc}$. This is due to the fact that in this regime the system is more similar to a sparse random graph rather than to a dense random matrix.

\end{thebibliography}
\end{document}